\documentclass[preprint2]{aastex}
\usepackage{graphicx}
\usepackage{graphics}
\newcommand{\beq}{\begin{eqnarray}}
\newcommand{\eeq}{\end{eqnarray}}
\newcommand{\bdm}{\begin{displaymath}}
\newcommand{\edm}{\end{displaymath}}
\newcommand{\be}{\begin{equation}}
\newcommand{\ee}{\end{equation}}

\newcommand{\pd}[1]{\, \partial #1 \,}
\newcommand{\td}[1]{\, \mathrm{d} #1 \,}
\newcommand{\intl}{\int\limits}

\newcommand{\HF}[1]{\; H\left[ #1 \right]}  
\newcommand{\DF}[1]{\; \delta\left( #1 \right)}  
\newcommand{\nup}[1]{\nu^{\prime #1}}
\newcommand{\SED}{f^{\prime}}
\newcommand{\ergcm}{\;\hbox{erg}\:\hbox{cm}^{-3}}
\newcommand{\ergs}{\;\hbox{erg}/\hbox{s}}
\newcommand{\nus}{\nu_{syn}}
\newcommand{\nut}{\nu_{T}}
\newcommand{\nub}{\nu_{B}}
\newcommand{\nug}{\nu_{\gamma_0}}
\newcommand{\nuc}{\nu_{c}}

\newcommand{\nubr}{\nu_{br}}
\newcommand{\nugbr}{\nu_{\gamma_{B}r}}
\newcommand{\nugr}{\nu_{\gamma_{0}r}}
\newcommand{\nugb}{\nu_{\gamma_B}}
\newcommand{\nucut}{\nu_{cut}}
\newcommand{\us}{\;\hbox{Hz}}

\newcommand{\Kl}{K\ll 1}
\newcommand{\Kg}{K\gg 1}
\newcommand{\Kla}{1\ll K\ll \alpha^3}
\newcommand{\Kga}{1\ll \alpha^3 \ll K}
\newcommand{\lec}{(1+l_{ec})}
\newcommand{\epss}{\epsilon_s}
\newcommand{\epec}{\epsilon_{ec}}
\newcommand{\sqrteeq}{\sqrt{1+\frac{1}{\epsilon_s\epsilon_{ec}q}}}
\newcommand{\ecut}{\epsilon_{cut}}
\newcommand{\epgb}{\epsilon_{\gamma_B}}
\newcommand{\sqrttwo}{2(\sqrt{2}-1)}
\received{}
\revised{}
\accepted{}
\shorttitle{External Compton radiation of SSC cooled electrons}
\shortauthors{M. Zacharias \& R. Schlickeiser}
\begin{document}
\title{External Compton emission in blazars of non-linear SSC cooled electrons}
\author{Michael Zacharias \& Reinhard Schlickeiser}
\affil{Institut f\"ur Theoretische Physik, Lehrstuhl IV: Weltraum- und Astrophysik, Ruhr-Universit\"at Bochum, 
44780 Bochum, Germany}
\email{mz@tp4.rub.de, rsch@tp4.rub.de}
\begin{abstract}
The origin of the high-energy component in spectral energy distributions (SED) of blazars is still a bit of a mystery. While BL Lac objects can be rather successfully modeled within the one-zone synchrotron self-Compton (SSC) scenario, {the SED of low peaked Flat Spectrum Radio Quasars (FSRQ) is more difficult to reproduce}. Their {high-energy component} needs the abundance of strong external photon sources, giving rise to stronger cooling via the inverse Compton channel, and thusly to a powerful component in the SED. Recently, we were able to show that such a powerful inverse Compton component can also be achieved within the SSC framework. This, however, {is only possible} if the electrons cool by SSC, which results in a non-linear process, since the cooling depends on an energy integral over the electrons. In this paper we aim to compare the non-linear SSC framework with the external Compton (EC) output by calculating analytically the external Compton {component} with the underlying electron distribution being either linearly or non-linearly cooled. Due to the additional linear cooling of the electrons with the external photons{, higher} number densities of electrons are required to achieve non-linear cooling, resulting in more powerful inverse Compton components. If the electrons initially cool non-linearly, the resulting SED {can exhibit a dominating SSC over the EC component. However, this dominance depends strongly on the input parameters. We conclude that with the correct time-dependent treatment the SSC component should be taken into account to model blazar flares.}
\end{abstract}
\keywords{radiation mechanisms: non-thermal -- BL Lacertae objects: general -- gamma-rays: theory}
%
%
%
%
\section{Introduction}
Combined as blazars, flat spectrum radio quasars (FSRQ) and BL Lacertae objects (BL Lacs) are the most {violent} subgroup of active galactic nuclei (AGN) {from the earth's point of view} in the accepted unification scheme (Urry \& Padovanni 1995). The broadband spectral energy distribution of blazars is dominated by two broad non-thermal {components}. The low-energetic one, peaking usually between the infrared and the X-ray parts, is attributed to synchrotron radiation of highly relativistic electrons, while the process behind the high-energy component, peaking in the $\gamma$-rays, is a matter of ongoing discussions. 

Albeit the possibility of a hadronic origin (e.g. Mannheim 1993) for the high-energetic {component}, most authors favor a leptonic origin of the $\gamma$-radiation (for recent reviews see B\"ottcher 2007, 2012). If highly relativistic electrons (and also positrons) interact with an ambient photon field, the photons can be inverse Compton scattered to very high energies. Such photon fields can be the synchrotron photons of the same population of electrons (the so-called synchrotron self-Compton effect (SSC), Jones, O'Dell \& Stein 1974), or photon sources external to the jet (so-called external Compton models (EC)), like photons directly from the accretion disk surrounding the black hole in the center of the active galaxy (Dermer \& Schlickeiser 1993), from the broad line regions (Sikora et al. 1994) or the dusty torus (Blazejowski et al. 2000, Arbeiter et al. 2002).

Especially the SEDs of FSRQs are dominated by the inverse Compton {component}, i.e. most of the luminosity of this type of blazars is emitted in $\gamma$-rays (e.g. Hayashida et al. 2012 for 3C 279, or Vercellone et al. 2011 for 3C 454.3). Since one can measure several thermal emission {components} in FSRQs, the EC process seems to be a natural choice to model the SED. Being able to detect {unbeamed thermal emission}, albeit the {strongly boosted} non-thermal radiation {of the jet, proves the abundance of} strong external photon fields. This in turn provides {lots} of seed photons for the electrons, which cool rather strongly by this process, giving rise to a powerful inverse Compton {component}.

On the other hand, BL Lacs emit most of their power in the synchrotron {component} having {at most comparable inverse Compton fluxes. Especially high-frequency peaked BL Lacs exhibit a much reduces $\gamma$-ray flux compared to the synchrotron flux.} This points towards a main cooling by the synchrotron channel, and the SSC process is successfully used to model BL Lacs (e.g. Acciari et al. 2011 for 1ES 2344+514, or Abramowski et al. 2012 for 1RXS J101015.9-311909).

However, such a strict division of the two blazar types has been called into question, recently. Chen et al. (2012) performed a numerical analysis of the multiwavelength variability of the FSRQ PKS 1510-089. Using a time-dependent code they were unable to find a clear preference of the EC over the SSC process, and state that the SSC process might even be preferable, since it matches the X-ray part of the SED far better than the EC process.

The advantage of the time-dependent numerical treatment over the usual steady-state approach is that such codes naturally implement the time-dependent nature of the SSC process, which is normally forgotten in many theoretical investigations (e.g. Moderski et al. 2005, Nakar et al. 2009) and modeling attempts (e.g. Ghisellini et al. 2009, Aleksic et al. 2012). Imagine an electron population that emits synchrotron radiation and then scatters this self-made radiation up to $\gamma$-rays. The electrons, therefore, lose energy, which means that the emitted synchrotron emission will also be less energetic, as will be the SSC emission. This implies that the cooling rate will also become weaker over time, i.e. the cooling rate is time-dependent. Schlickeiser (2009) gave an analytical expression for the time-dependent SSC cooling rate, which is proportional to an energy integral over the electron distribution function itself.

Schlickeiser, B\"ottcher \& Menzler (2010, hereafter referred to as SBM) combined the new SSC cooling term with the well known synchrotron cooling term, in order to give a more realistic treatment. As mentioned before, the SSC cooling becomes weaker over time, which means that after some time it will be weaker than the standard linear cooling terms, such as synchrotron cooling, since they do not depend on time. They also calculated the synchrotron SED with the remarkable result that the synchrotron {component} exhibits a broken power-law behaviour. Interestingly, this feature is independent of the electron injection function. While SBM used a $\delta$-like injection, Zacharias \& Schlickeiser (2010) performed the same analytical calculation using a power-law injection. Apart from the high-energy end of the synchrotron {component}	the broken power-law behaviour is the same in both cases. This is due to the fact that any extended form of the electron distribution is quickly quenched into a $\delta$-like structure, as long as no reacceleration is taken into account. Therefore, the $\delta$-approach is sort of a late time limit for any extended injection.

Following-up on the results described above, Zacharias \& Schlickeiser (2012, hereafter referred to as ZS) calculated analytically for the $\delta$-approach the emerging SSC SED. They obtained the interesting result that the time-dependent SSC cooling leads, in fact, to a dominating inverse Compton (IC) {component} without the need for rather extreme parameter settings. Additionally, they found that the SSC {component} also exhibits a broken power-law, which may also be independent of the injection form.\footnote{We note that the type of breaks in the SED that we can naturally account for with our model, are usually described with rather complicated electron distributions, {requiring sometimes multiple spectral breaks in the electron source energy distribution} with practically no theoretical justification (e.g. Abdo et al. 2011a for Mrk 501). }

As we already discussed above, the debate whether SSC, EC or both play an important role especially in FSRQs is not yet settled. Since we showed in ZS that a dominating SSC {component} is easily achievable, it is just straightforward to include the EC scenario in our approach. Therefore, we will extend the loss rate of SBM with the additional contribution of the external Compton losses. This will be done in section \ref{sec:lr}. {In section \ref{sec:flu} we will calculate the resulting intensity and fluence spectrum of the EC {component}, where the fluence is the time-integrated, i.e. averaged, intensity. The lengthy details of these calculations can be found in appendix \ref{sec:app2} and \ref{sec:app3}.} Transforming the results into the form of an SED will be done in section \ref{sec:sed}, where we will also summarise the results of ZS for the sake of completeness. At the end of this section we will give some example SEDs with all contributions of synchrotron, SSC and EC, and will discuss the results. In section \ref{sec:con} we will summarise the results and present our conclusions.

%
\section{The extended loss rate} \label{sec:lr}

Since we intend to calculate the spectrum due to external Compton emission, we have to include this type of energy transfer between electrons and photons in the energy loss rate of the relativistic electrons. According to Dermer \& Schlickeiser (1993) the pitch-angle averaged loss rate of electrons in an external photon field that is isotropically distributed in the lab frame, is
\beq
|\dot{\gamma}_{ec}| &=& \frac{4c\sigma_Tu^{\prime}_{ec}\gamma^2}{3m_ec^2}\Gamma_b^2\left( \frac{4}{3} - \frac{1}{3\Gamma_b^2} \right) \nonumber \\
&\approx& \frac{16c\sigma_Tu^{\prime}_{ec}\Gamma_b^2}{9m_ec^2}\gamma^2 , 
\label{exlossrate}
\eeq
with the speed of light $c=3\times 10^{10}\,\mathrm{cm/s}$, the Thomson cross section $\sigma_T=6.65\times 10^{-25}\,\mathrm{cm^2}$, the energy of an electron at rest $m_ec^2=8.14\times 10^{-7}\,\mathrm{erg}$, the electron Lorentz factor $\gamma$, and the Lorentz factor of the radiating plasma blob $\Gamma_b$.

We assume that the energy density of the external photons is isotropic in the galactic {(primed)} frame, and is given by
\be
u^{\prime}_{ec} = \frac{L^{\prime}_{ad}\tau_{sc}}{4\pi R_{sc}^{\prime 2}c} ,
\label{exenergydensity}
\ee
where $L^{\prime}_{ad}$ is the luminosity of the accretion disk surrounding the central supermassive black hole, $\tau_{sc}$ is the scattering depth of the ambient medium scattering the accretion disk photons, and $R_{sc}^{\prime}$ is the radius up to where the accretion disk photons are scattered by the ambient medium.

Adding the external cooling term to the synchrotron and SSC cooling terms, we obtain the complete electron loss rate:
\beq
|\dot{\gamma}_{tot}| &=& |\dot{\gamma}_{syn}|+|\dot{\gamma}_{ec}|+|\dot{\gamma}_{ssc}| \nonumber \\
&=& D_0\gamma^2 + \frac{16c\sigma_Tu^{\prime}_{ec}\Gamma_b^2}{9m_ec^2}\gamma^2 + A_0\gamma^2 \intl_0^{\infty}\td{\gamma} \gamma^2 n(\gamma,t) \nonumber \\
&=& D_0 \left( 1+l_{ec} \right) \gamma^2 + A_0\gamma^2 \intl_0^{\infty}\td{\gamma} \gamma^2 n(\gamma,t) ,
\label{totcoolrate}
\eeq
with $D_0 = 4c\sigma_Tu_B/(3m_ec^2) = 1.29\times 10^{-9}b^2\,\mathrm{s^{-1}}$, the magnetic energy density $u_B = B^2/8\pi$, {and the magnetic field $B = b\,\mathrm{Gauss}$. Schlickeiser (2009) gives the constant $A_0 = 3\sigma_Tc_1P_0R\epsilon^2_{0}/(m_ec^2) = 1.15\times 10^{-18}R_{15}b^2\,\mathrm{cm^3s^{-1}}$, which was obtained during the derivation of the time-dependent SSC cooling term. The parameters are given as $P_0 = 2\times 10^{24}\,\mathrm{erg^{-1}s^{-1}}$, $\epsilon_{0} = 1.856\times 10^{-20}b\,\mathrm{erg}$, and $c_1=0.684$. We scaled the radius of the spherical emission blob as $R=10^{15}R_{15}\,\mathrm{cm}$.}

The non-linearity of the SSC cooling manifests itself in the integral over the volume-averaged electron distribution $n(\gamma,t)$. The factor
\beq
l_{ec} &=& \frac{u^{\prime}_{ec}}{u_B}\frac{4\Gamma_b^2}{3} = \frac{8L^{\prime}_{ad}\tau_{sc}\Gamma_b^2}{3B^2R_{sc}^{\prime 2}c} \nonumber \\
&=& 0.09 \frac{L^{\prime}_{46}\tau_{-2}\Gamma_{b,1}^2}{b^2R_{pc}^{\prime 2}} 
\label{externalratio}
\eeq
shows the relative strength of external to synchrotron cooling. For $l_{ec}\gg 1$ the linear cooling is dominated by the external photons, while for $l_{ec}\ll 1$ the synchrotron process mainly operates. Apart from $R_{sc}^{\prime}$, which is scaled as $R_{sc}^{\prime} = 3.08\times 10^{18} R_{pc}^{\prime}\,\mathrm{cm}$, we scaled the quantities in equation (\ref{externalratio}) as $Q = 10^x Q_x$ in their respective cgs-units{, and $\Gamma_{b}=10\Gamma_{b,1}$}.

We can now formulate the differential equation describing the competition between the injection of ultrarelativistic particles with the source function $S(\gamma,t)$ and the energy losses as described by equation (\ref{totcoolrate}) inside the spherical emission region (Kardashev 1962):
\be
\frac{\pd{n(\gamma,t)}}{\pd{t}} - \frac{\pd}{\pd{\gamma}}\left[ |\dot{\gamma}_{tot}|n(\gamma,t) \right] = S(\gamma,t) .
\label{PDEn1}
\ee
In order to keep the problem simple, we assume a monoenergetic instantaneous injection $S(\gamma,t) = q_0\DF{\gamma-\gamma_0}\DF{t}$. Inserting equation (\ref{totcoolrate}) into equation (\ref{PDEn1}) we obtain
\beq
\frac{\pd{n}}{\pd{t}} - \frac{\pd}{\pd{\gamma}}\left\{ \left[D_0 \left( 1+l_{ec} \right) + A_0 \intl_0^{\infty}\td{\gamma} \gamma^2 n(\gamma,t)\right] \gamma^2 n \right\} \nonumber \\
= q_0\DF{\gamma-\gamma_0}\DF{t} ,
\label{PDEn2}
\eeq
which apart from the factor $(1+l_{ec})$ equals the differential equation that was solved by SBM. We can, therefore, use their solution. The only difference is that we have to insert the factor $\lec$ wherever they have a $D_0$. {The solution is given in cases of the injection parameter $\alpha$, which is proportional to the ratio of the nonlinear to linear cooling at time of injection. We will discuss its implications in section \ref{sec:injpara}.}

For dominating linear cooling $(\alpha\ll 1)$, we get the electron distribution
\be
n(\gamma,t,\alpha\ll 1) = q_0\DF{\gamma-\frac{\gamma_0}{1+D_0\lec\gamma_0t}} .
\label{nlinsol}
\ee

If initially the non-linear cooling dominates $(\alpha\gg 1)$, we obtain
\beq
n(\gamma,t,\alpha\gg 1) = q_0 \HF{t_c-t} \nonumber \\
\times \DF{\gamma-\frac{\gamma_0}{(1+3D_0\lec\gamma_0\alpha^2 t)^{1/3}}} ,
\label{nnonlinsol1}
\eeq
which is valid for times 
\be
t\leq t_c = \frac{\alpha^3-1}{3D_0\lec\gamma_0\alpha^2} .
\label{tc}
\ee
For late times the linear cooling takes over and the distribution function is described by a modified linear solution:
\beq
n(\gamma,t,\alpha\gg 1) = q_0 \HF{t-t_c} \nonumber \\
\times \DF{\gamma-\frac{\gamma_0}{\frac{1+2\alpha^3}{3\alpha^2}+D_0\lec\gamma_0 t}} .
\label{nnonlinsol2}
\eeq


\subsection{The injection parameter $\alpha$} \label{sec:injpara}

We have written the solutions of the differential equation (\ref{PDEn2}) dependent on the parameter $\alpha$, which we intend to discuss in greater detail, now.

We call it the injection parameter, and it is defined as the square-root of the ratio of the non-linear cooling term to the linear cooling term at time of injection, i.e.
\beq
\alpha^2 &=& \frac{|\dot{\gamma}_{ssc}(t=0)|}{|\dot{\gamma}_{syn}|+|\dot{\gamma}_{ec}|} = \frac{A_0q_0\gamma_0^2}{D_0\lec} \nonumber \\
&=& 2.13\times 10^{3} \frac{N_{50}\gamma_4^2}{R_{15}^2 \lec},
\label{alpha1}
\eeq
{where $N$ denotes the number of radiating particles, and we applied the same scaling law for the parameters as above.}

For $\alpha\ll 1$ we see that the linear cooling dominates, resulting in the solution (\ref{nlinsol}). If $\alpha\gg 1$, the non-linear cooling at least initially dominates, giving the solutions (\ref{nnonlinsol1}) and (\ref{nnonlinsol2}) for early and late times, respectively. The time $t_c$ marks the transition from non-linear to linear cooling.

Comparing the above given $\alpha$ to the injection parameter obtained by SBM (where EC losses were neglected), we find that
\be
\alpha = \frac{\alpha_{SBM}}{\lec^{1/2}} ,
\label{alphaSBM}
\ee
where $\alpha_{SBM} = A_0q_0\gamma_0^2/D_0$.

This implies that the contribution of the external photons lowers the possibility for non-linear cooling. Solving equation (\ref{alpha1}) for the electron density we obtain
\be
q_0 = \frac{D_0\lec}{A_0\gamma_0^2}\alpha^2 > \frac{D_0}{A_0\gamma_0^2}\alpha_{SBM}^2 = q_{0,SBM} .
\label{q0eldens}
\ee
This also demonstrates the afore mentioned fact that with external Compton losses included it is harder to cool the electrons non-linearly. Equation (\ref{q0eldens}) shows that for the same value of $\alpha$, i.e. the relative strength between linear and non-linear cooling, one needs a higher electron density in the blob compared to the case where the external losses are neglected. 

%
%
\section{External Compton fluence} \label{sec:flu}

The intensity due to inverse Compton collisions of electrons with external photons is given by
\be
I_{ec}(\epsilon_s,t) = Rj_{ec}(\epsilon_s,t) ,
\label{ecintens}
\ee
with the emissivity (see appendix \ref{sec:app1})
\be
j_{ec}(\epsilon_s,t) = \frac{c\epsilon_s}{4\pi} \intl_0^{\infty}\td{\epsilon}\frac{u(\epsilon)}{\epsilon} \intl_1^{\infty}\td{\gamma}n(\gamma,t)\sigma(\epsilon_s,\epsilon,\gamma) ,
\label{ecemi1}
\ee
where $\epsilon$ is the normalized target photon energy {in units of the electron rest mass $m_ec^2$}, $\epsilon_s$ is the normalized scattered photon energy, $u(\epsilon)$ is the target photon density, and 
\be
\sigma(\epsilon_s,\epsilon,\gamma) = \frac{3\sigma_T}{4\epsilon\gamma^2}G(q,\Gamma) 
\label{kncs1}
\ee
being the Klein-Nishina cross-section (Blumenthal \& Gould 1970) with
\beq
G(q,\Gamma) &=& G_0(q)+\frac{\Gamma^2q^2(1-q)}{2(1+\Gamma q)} , \\
G_0(q) &=& 2q\ln{q}+1+q-2q^2 , \\
\Gamma &=& 4\epsilon\gamma , \\
q &=& \frac{\epsilon_s}{\Gamma(\gamma-\epsilon_s)} .
\eeq
{ Thus, we obtain for equation (\ref{ecemi1})
\be
j_{ec}(\epsilon_s,t) = \frac{3c\sigma_T}{16\pi}\epsilon_s \intl_0^{\infty}\td{\epsilon}\frac{u(\epsilon)}{\epsilon^2} \intl_{\gamma_{min}}^{\infty}\td{\gamma}\frac{n(\gamma,t)}{\gamma^2}G(q,\Gamma) .
\label{ecemi2}
\ee}
Here
\be
\gamma_{min}(\epsilon_s,\epsilon) = \frac{\epsilon_s}{2}\left[ 1+\sqrt{1+\frac{1}{\epsilon\epsilon_s}} \right]
\label{gammin}
\ee
denotes the minimum Lorentz factor for the electrons, below which the electrons would gain energy from the photons.

{The fluence is the time integrated intensity spectrum, giving an average of the variability in all bands, and also incorporating that observation times can be much longer than the typical flare duration:
\be
F(\epsilon_s) = \intl_{0}^{\infty}\td{t}I(\epsilon_s,t) .
\label{fluall}
\ee }

The intention of this paper is to calculate analytically the complete SED of blazars. { We will therefore stick to a rather simple approach and use the simplest approach possible for the external photon density in the comoving frame:
\be
u(\epsilon) = \frac{4}{3}\Gamma_b^2 u^{\prime}_{ec}\DF{\epsilon-\epsilon_{ec}} ,
\label{exphotden}
\ee
where $\epsilon_{ec}$ is the normalized energy of the target photons in the comoving frame.} Using the electron densities (\ref{nlinsol}), (\ref{nnonlinsol1}), and (\ref{nnonlinsol2}) we can calculate the intensity, and afterwards the fluence. 


\subsection{Small injection parameter, $\alpha\ll 1$}

{Using equation (\ref{nlinsol}) in equation (\ref{ecintens}) we find for the case of $\alpha\ll 1$:
\beq
I_{ec}(\epsilon_s,\tau) = I_0\epsilon_s (1+\tau)^2 G\left( q(\tau),\Gamma(\tau),\epsilon=\epsilon_{ec},\gamma=\frac{\gamma_0}{1+\tau} \right) \nonumber \\
 \times \HF{\tau} \HF{\gamma_0-\gamma_{min}(\epsilon=\epsilon_{ec})(1+\tau)} ,
\label{linintens2}
\eeq
after performing the simple integrations of the $\delta$-functions, and substituting $\tau=D_0\lec\gamma_0t$. { Here we defined $I_0=Rc\sigma_Tq_0\Gamma_b^2u^{\prime}_{ec}/(4\pi\epsilon_{ec}^2\gamma_0^2)$, }
\beq
G\left( q(\tau),\Gamma(\tau),\epsilon=\epsilon_{ec},\gamma=\frac{\gamma_0}{1+\tau} \right)  \nonumber \\ 
= G_0(q)+\frac{\epsilon_s^2(1-q)(1+\tau)^2}{2\gamma_0(\gamma_0-\epsilon_s(1+\tau))} , 
\label{Glin1}
\eeq
\beq
\Gamma(\tau) &=& 4\epsilon_{ec}\frac{\gamma_0}{1+\tau} , \\
q(\tau) &=& \frac{\epsilon_s(1+\tau)^2}{4\epsilon_{ec}\gamma_0(\gamma_0-\epsilon_s(1+\tau))} , \\
\gamma_{min}(\epsilon=\epsilon_{ec}) &=& \frac{\epsilon_s}{2}\left[ 1+\sqrt{1+\frac{1}{\epsilon_{ec}\epsilon_s}} \right] .
\label{Glin2all}
\eeq


Inserting equation (\ref{linintens2}) into equation (\ref{fluall}) we obtain after some calculations (see appendix \ref{sec:app2}) the final expression for the external Compton fluence in the case of dominating linear cooling:
\be
F(\epss) = F_0\gamma_0^3\epec^{3/2} \frac{\epss^{-1/2}}{(1+\epec\epss)^{3/2}} \left( 1-\frac{\epss}{\ecut} \right) ,
\label{flulinsolges}
\ee
where $F_0 = I_0/(D_0\lec\gamma_0)$, and $\ecut = 4\epec\gamma_0^2/(1+4\epec\gamma_0)$.

{ In the Thomson-limit $\Gamma_0=4\epec\gamma_0\ll 1$ the spectrum cuts off at $\epss\approx \Gamma_0\gamma_0\ll \gamma_0$. The Thomson-limit also implies that $\epec\ll (4\gamma_0)^{-1}$. Hence, there is no break in the spectrum, and the denominator in equation (\ref{flulinsolges}) equals unity. 

In the Klein-Nishina-limit $\Gamma_0\gg 1$ we find $\epec\gg (4\gamma_0)^{-1}$, and the cut-off at $\epss\approx \gamma_0$. Thus, the spectrum breaks at $\epss=\epec^{-1}$.}


\subsection{Large injection parameter, $\alpha\gg 1$}


\subsubsection{Early time limit, $t\leq t_c$}

{For $t\leq t_c$ we use equation (\ref{nnonlinsol1}) in equation (\ref{ecintens}), as well as the same substitution for $t$ as above, and obtain after solving the simple integrations
\beq
I_{ec}(\epsilon_s,\tau) = I_0\epsilon_s \HF{\tau_c-\tau}(1+3\alpha^2\tau)^{2/3} \nonumber \\
\times G\left( q(\tau),\Gamma(\tau),\epsilon=\epsilon_{ec},\gamma=\frac{\gamma_0}{(1+3\alpha^2\tau)^{1/3}} \right) \nonumber \\
 \times \HF{\tau} \HF{\gamma_0-\gamma_{min}(\epsilon=\epsilon_{ec})(1+3\alpha^2\tau)^{1/3}} ,
\label{nonlinintens12}
\eeq
with $\tau_c = (\alpha^3-1)/3\alpha^2$, 
\beq
G\left( q(\tau),\Gamma(\tau),\epsilon=\epsilon_{ec},\gamma=\frac{\gamma_0}{(1+3\alpha^2\tau)^{1/3}} \right)  \nonumber \\ 
= G_0(q)+\frac{\epsilon_s^2(1-q)(1+3\alpha^2\tau)^{2/3}}{2\gamma_0 \left( \gamma_0-\epsilon_s(1+3\alpha^2\tau)^{1/3} \right)} , 
\label{Gnonlin1}
\eeq
\beq
\Gamma(\tau) &=& 4\epsilon_{ec}\frac{\gamma_0}{(1+3\alpha^2\tau)^{1/3}} , \\
q(\tau) &=& \frac{\epsilon_s(1+3\alpha^2\tau)^{2/3}}{4\epsilon_{ec}\gamma_0 \left( \gamma_0-\epsilon_s(1+3\alpha^2\tau)^{1/3} \right)} .
\eeq


Using equation (\ref{fluall}) with equation (\ref{nonlinintens12}) the fluence for the early time limit can be calculated. The lengthy details can be found in appendix \ref{sec:app3}. The solution depends on the external photon energy $\epec$, giving three different cases. \\
For $\epec^{-1}<\sqrttwo\gamma_B$ we have
\beq
F(\epss) = \frac{F_0\alpha^3}{5} \frac{\epss}{\left( 1+\frac{\epss}{\epgb} \right)\left( 1+\frac{\epss}{\sqrttwo\gamma_B} \right)^{4}} \nonumber \\
\times \left( 1-\frac{\epss}{\ecut} \right) ,
\label{flunonling1}
\eeq
while for $\sqrttwo\gamma_B<\epec^{-1}<\sqrttwo\gamma_0$ we get
\beq
F(\epss) = \frac{F_0\alpha^3}{5} \frac{\epss}{\left( 1+\frac{\epss}{\epgb} \right)^{5/2} \left( 1+\epec\epss \right)^{5/2}} \nonumber \\
\times \left( 1-\frac{\epss}{\ecut} \right) . 
\label{flunonling2}
\eeq
The last part is for $\sqrttwo\gamma_0<\epec^{-1}$ and becomes
\beq
F(\epss) = \frac{F_0\alpha^3}{5} \frac{\epss}{\left( 1+\frac{\epss}{\epgb} \right)^{5/2} \left( 1+\frac{\epss}{\sqrttwo\gamma_B} \right)^{5/2}} \nonumber \\
\times \left( 1-\frac{\epss}{\ecut} \right) . 
\label{flunonling3}
\eeq
Here we used $\gamma_B=\gamma_0/\alpha$, { and $\epgb = 4\epec\gamma_B^2/(1+4\epec\gamma_B)$.}


\subsubsection{Late time limit, $t\geq t_c$}

As before, we can find the intensity for $t\geq t_c$, if we insert equation (\ref{nnonlinsol2}) into equation (\ref{ecintens}). As in the previous cases we exchange $t$, and get
\beq
I_{ec}(\epsilon_s,\tau) = I_0\epsilon_s\HF{\tau-\tau_c} (\frac{1+2\alpha^3}{3\alpha^2}+\tau)^2 \nonumber \\
\times G\left( q(\tau),\Gamma(\tau),\epsilon=\epsilon_{ec},\gamma=\frac{\gamma_0}{\frac{1+2\alpha^3}{3\alpha^2}+\tau} \right) \nonumber \\
 \times \HF{\gamma_0-\gamma_{min}(\epsilon=\epsilon_{ec})(\frac{1+2\alpha^3}{3\alpha^2}+\tau)} ,
\label{nonlinintens22}
\eeq
with 
\beq
G\left( q(\tau),\Gamma(\tau),\epsilon=\epsilon_{ec},\gamma=\frac{\gamma_0}{1+\tau} \right)  \nonumber \\ 
= G_0(q)+\frac{\epsilon_s^2(1-q)(\frac{1+2\alpha^3}{3\alpha^2}+\tau)^2}{2\gamma_0 \left( \gamma_0-\epsilon_s(\frac{1+2\alpha^3}{3\alpha^2}+\tau) \right)} , 
\label{Gnonlin2}
\eeq
\beq
\Gamma(\tau) &=& 4\epsilon_{ec}\frac{\gamma_0}{\frac{1+2\alpha^3}{3\alpha^2}+\tau} , \\
q(\tau) &=& \frac{\epsilon_s(1+\tau)^2}{4\epsilon_{ec}\gamma_0(\gamma_0-\epsilon_s(1+\tau))} .
\eeq


Inserting equation (\ref{nonlinintens22}) into equation (\ref{fluall}) one can obtain the fluence in the late time limit. For the details we refer the reader to appendix \ref{sec:app3}, again. The results depend also on the external photon energy, giving two cases this time. \\
For $\epec^{-1}<\sqrttwo\gamma_B$ we find
\beq
F(\epss<\epgb) = F_0\gamma_0^3\epec^{3/2} \frac{\epss^{-1/2}}{(1+\epec\epss)^{3/2}} \left( 1-\frac{\epss}{\epgb} \right) ,
\label{flunonling4}
\eeq
while for $\epec^{-1}>\sqrttwo\gamma_B$ we obtain a single power-law in the form
\beq
F(\epss<\epgb) = \frac{16}{3}F_0\epec^{3/2}\gamma_0^3\epss^{-1/2} \left( 1-\frac{\epss}{\epgb} \right) .
\label{flunonling5}
\eeq
}


\subsubsection{Total fluence in the case $\alpha\gg 1$}

Combining the results of the previous sections we obtain the total fluence for the inverse Compton {component} due to interactions with the ambient radiation field in the case that the electrons are {at first} cooled non-linearly. We have three different cases depending on the value of the normalized external photon energy $\epec$. \\
For $\epec^{-1}<\sqrttwo\gamma_B$ we find
\beq
F(\epss) = F_0\gamma_0^3\epec^{3/2} \frac{\epss^{-1/2}}{\left( 1+\epec\epss \right)^{3/2} \left( 1+\frac{\epss}{\epgb} \right)^2} \nonumber \\
\times \left( 1-\frac{\epss}{\ecut} \right) .
\label{flunonlinG1}
\eeq
Secondly, if $\sqrttwo\gamma_B<\epec^{-1}<\sqrttwo\gamma_0$ the fluence becomes
\beq
F(\epss) = \frac{16}{3}F_0\gamma_0^3\epec^{3/2} \frac{\epss^{-1/2}}{\left( 1+\frac{\epss}{\epgb} \right)  \left( 1+\epec\epss \right)^{5/2}} \nonumber \\
\times \left( 1-\frac{\epss}{\ecut} \right) .
\label{flunonlinG2}
\eeq
Lastly, we obtain for $\sqrttwo\gamma_0<\epec^{-1}$
\beq
F(\epss) = \frac{16}{3}F_0\gamma_0^3\epec^{3/2} \frac{\epss^{-1/2}}{\left( 1+\frac{\epss}{\epgb} \right)  \left( 1+\frac{\epss}{\sqrttwo\gamma_0} \right)^{5/2}} \nonumber \\
\times \left( 1-\frac{\epss}{\ecut} \right) .
\label{flunonlinG3}
\eeq
These are all broken power-laws, where at least one break ($\epgb$) depends strongly on the value of $\alpha$, and therefore on the non-linear cooling.

{ We note that the first case corresponds to the extreme Klein-Nishina-limit ($\Gamma_0\gg 1$), the second one to the mild Klein-Nishina-limit ($\Gamma_0>1$), and the last one to the Thomson-limit ($\Gamma_0\lesssim 1$).}
%
%
\section{The complete SED} \label{sec:sed}

Using the results of the last section and of ZS\footnote{Some of the results of ZS have printing errors, which we correct here.} we are now in a position to present the complete spectral energy distribution in a combined picture of SSC end EC radiation. The inverse Compton {component} will, therefore, be the sum of the SSC and EC contributions. If these contributions are comparable on some scales the resulting spectrum will deviate from pure power-laws.	

In order to give the results in a manner that can be easily compared to data, we will give the SEDs depending on the frequency $\nup{}=\frac{mc^2}{h}\delta\epsilon$, which is already transformed to the frame of rest of the host galaxy with the Doppler factor $\delta=(\Gamma_b(1-\beta_{\Gamma_b}\mu_{obs}))^{-1}$, where $\mu_{obs}$ is the cosine of the angle between the jet and the line of sight, {and $\beta_{\Gamma_b}=v_b/c$ is the normalized speed of the plasma blob}. The SED is then {given by the fluence multiplied with the frequency $\SED=\pi cR\nup{}F^{\prime}(\nup{})$ in units of $\ergs$} with the transformed fluence $F^{\prime} = \delta^4 F$. 

We should note that we have to adapt the results of ZS to the case discussed here. That is, we have to include the addition of the external Compton cooling to the linear term. This can be done by replacing any "$D_0$" of ZS with "$D_0\lec$". Interestingly, the synchrotron {component} will not be affected by this replacement, while the SSC {component} gains a factor $\lec$.

Below we will first present the theoretical SEDs and afterwards give a brief discussion of the results.
%
%
\subsection{Synchrotron SED}
From equation (ZS-69) we obtain the synchrotron SED for the case $\alpha\ll 1$, which is
\be
\SED_s(\nup{}) = 5.6 \cdot 10^{38} R_{15} \delta^4 \frac{\alpha^2}{\gamma_4} \left( \frac{\nup{}}{\nu_{syn}} \right)^{1/2} e^{-\nup{}/\nu_{syn}} \ergs ,
\label{sedsyn1}
\ee
with $\nu_{syn}=4.1\cdot 10^{14}\delta b\gamma_4^2\us$. \\
The maximum value of the synchrotron SED, 
\be
\SED_{s,max}=2.4\cdot 10^{38} R_{15} \delta^4 \frac{\alpha^2}{\gamma_4} \ergs,
\label{sedsyn1max}
\ee
is attained at
\be
\nup{}_{s,max}=\frac{1}{2}\nus = 2.1\cdot 10^{14}\delta b\gamma_4^2 \us. 
\label{sedsyn1fre}
\ee
Using equation (ZS-82) we can write the synchrotron SED for $\alpha\gg 1$ as
\be
\SED_{s} = 5.6\cdot 10^{38} R_{15} \delta^4 \frac{\alpha^2}{\gamma_4} \frac{\left( \frac{\nup{}}{\nus} \right)^{1/2}}{1+\frac{\nup{}}{\nuc}} e^{-\nup{}/\nus} \ergs .
\label{sedsyn2}
\ee
It peaks at
\be
\nup{}_{s,max} = \nuc = 2.9\cdot 10^{14} \delta \frac{b\gamma_4^2}{\alpha^2} \us
\label{sedsyn2fre}
\ee
with the maximum value
\be
\SED_{s,max} = 2.4\cdot 10^{38} R_{15} \delta^4 \frac{\alpha}{\gamma_4} \ergs .
\label{sedsyn2max}
\ee
%
%
\subsection{Synchrotron self-Compton SED}
For $\alpha\ll 1$ ZS found two different versions of the SSC SED depending on the Klein-Nishina parameter $K=0.136 b\gamma_4^3$. In the Thomson limit ($K\ll 1$) we use equation (ZS-72) and get
\beq
\SED_{SSC}(K\ll 1) = 2.8\cdot 10^{39} R_{15} \delta^4 \lec \frac{\alpha^4}{\gamma_4} \nonumber \\
\times \left( \frac{\nup{}}{\nut} \right)^{3/4} e^{-\nup{}/\nut} \ergs ,
\label{sedssc1}
\eeq
while for the Klein-Nishina limit ($K\gg 1$) with equation (ZS-73) we have
\beq
\SED_{SSC}(K\gg 1) = 2.1\cdot 10^{40} R_{15} \delta^4 \lec \frac{\alpha^4}{b\gamma_4^4} \nonumber \\
\times \frac{\left( \frac{\nup{}}{\nub} \right)^{3/4}}{\left( 1+\frac{\nup{}}{\nub} \right)^{7/4}}  \left[ 1-\left( \frac{\nup{}}{\nug} \right)^{7/12} \right] \ergs,
\label{sedssc2}
\eeq
with the constants $\nut=1.6\cdot 10^{23}\delta b\gamma_4^4\us$, the break frequency $\nub=2.3\cdot 10^{24}\delta b^{-1/3}\us$ (Schlickeiser \& R\"oken 2008), and $\nug=1.2\cdot 10^{24}\delta\gamma_4\us$. \\
The maximum frequencies
\be
\nup{}_{SSC,max}(K\ll 1) = \frac{3}{4}\nut = 1.2\cdot 10^{23} \delta b\gamma_4^4 \us ,
\label{sedssc1fre}
\ee
and
\be
\nup{}_{SSC,max}(K\gg 1) = \frac{3}{4}\nub = 1.7\cdot 10^{24} \delta b^{-1/3} \us
\label{sedssc2fre}
\ee
imply the maximum values
\be
\SED_{SSC,max}(\Kl) = 1.0\cdot 10^{39} R_{15} \delta^4 \lec \frac{\alpha^4}{\gamma_4} \ergs ,
\label{sedssc1max}
\ee
and
\be
\SED_{SSC,max}(\Kg) = 6.3\cdot 10^{39} R_{15} \delta^4 \lec \frac{\alpha^4}{b\gamma_4^4} \ergs ,
\label{sedssc2max}
\ee
respectively. \\

For the SSC SEDs in the case $\alpha\gg 1$ we have to discuss three different cases depending on the Klein-Nishina parameter $K$, and we begin with the Thomson limit ($\Kl$) using equation (ZS-85):
\beq
\SED_{SSC}(\Kl) = 2.8\cdot 10^{39} R_{15} \delta^{4} \lec \frac{\alpha^4}{\gamma_4} \nonumber \\
\times \frac{\left( \frac{\nup{}}{\nut} \right)^{3/4}}{\left( 1 + \frac{\alpha^4\nup{}}{\nut} \right)^{1/2}} e^{-\nup{}/\nut} \ergs .
\label{sedssc3}
\eeq
The maximum value 
\be
\SED_{SSC,max}(\Kl) = 1.6\cdot 10^{39} R_{15} \delta^4 \lec \frac{\alpha^2}{\gamma_4} \ergs
\label{sedssc3max}
\ee
is attained at
\be
\nup{}_{SSC,max}(\Kl) = \frac{1}{4}\nut = 4.0\cdot 10^{22} \delta b \gamma_4^4 \us .
\label{sedssc3fre}
\ee
For the mild Klein-Nishina limit ($\Kla$) equation (ZS-88) yields
\beq
\SED_{SSC}(\Kla) = 1.5\cdot 10^{39} R_{15} \delta^4 \lec \frac{\alpha^4}{\gamma_4} \nonumber \\
\times \frac{\left( \frac{\nup{}}{\nut} \right)^{3/4}}{\left( 1+\frac{\alpha^4\nup{}}{\nut} \right)^{1/2} \left( 1+\frac{\nup{}}{\nub} \right)^{13/4}} \left[ 1-\left( \frac{\nup{}}{\nug} \right)^{13/3} \right] \ergs .
\label{sedssc4}
\eeq
This triple power-law peaks at
\be
\nup{}_{SSC,max}(\Kla) \approx \frac{1}{12}\nub = 1.9\cdot 10^{23} \delta b^{-1/3} \us ,
\label{sedssc4fre}
\ee
and reaches a maximum value of
\be
\SED_{SSC,max}(\Kla) \approx 1.2\cdot 10^{39} R_{15} \delta^4 \lec \frac{\alpha^2}{b^{1/3}\gamma_4^2} \ergs .
\label{sedssc4max}
\ee
The last case is the extreme Klein-Nishina limit ($\Kga$). We obtain the SED from equation (ZS-91), resulting in
\beq
\SED_{SSC}(\Kga) = 1.1\cdot 10^{40} R_{15} \delta^4 \lec \frac{\alpha^4}{b\gamma_4^4} \nonumber \\
\times  \frac{\left( \frac{\nup{}}{\nub} \right)^{3/4}}{\left( 1+\frac{\nup{}}{\nub} \right)^{7/4} \left( 1+\frac{\alpha\nup{}}{\nug} \right)^{2}} \left[ 1-\left( \frac{\nup{}}{\nug} \right)^{13/3} \right] \ergs .
\label{sedssc5}
\eeq
The SED peaks with a maximum value
\be
\SED_{SSC,max}(\Kga) \approx 3.4\cdot 10^{39} R_{15} \delta^4 \lec \frac{\alpha^4}{b \gamma_4^4} \ergs
\label{sedssc5max}
\ee
at a peak frequency
\be
\nup{}_{SSC,max}(\Kga) \approx \frac{3}{4}\nub = 1.7\cdot 10^{24} \delta b^{-1/3} \us .
\label{sedssc5fre}
\ee

%
%
%
\subsection{External Compton SED}
{ In the case of $\alpha\ll 1$ the SED can be calculated from equation (\ref{flulinsolges}), becoming
\beq
\SED_{ec}(\nup{}) = 4.0\times 10^{34} R_{15} l_{ec}\frac{\delta^4\alpha^2}{\gamma_4^2\epec} \nonumber \\
\times \frac{\left( \frac{\nup{}}{\nubr} \right)^{1/2}}{\left( 1+\frac{\nup{}}{\nubr} \right)^{3/2}} \left( 1-\frac{\nup{}}{\nucut} \right) \ergs ,
\label{sedec0}
\eeq
where $\nubr=1.23\cdot 10^{20} \delta \epec^{-1} \us$, and $\nucut=1.23\cdot 10^{20}\delta \ecut \us$. In the Thomson-limit ($\epec\ll (4\gamma_0)^{-1}$) the maximum frequency 
\beq
\nup{}_{ec,max} \approx 1.6\times 10^{20}\delta\epec\gamma_0^2 \us 
\label{sedec01fre}
\eeq
leads to a peak value of
\beq
\SED_{ec,max} \approx 4.4\times 10^{38} R_{15} l_{ec}\frac{\delta^4\alpha^2}{\gamma_4} \ergs .
\label{sedec01max}
\eeq
In the Klein-Nishina-limit ($\epec\gg (4\gamma_0)^{-1}$) the maximum value 
\beq
\SED_{ec,max} \approx 1.6\times 10^{34} R_{15} l_{ec}\frac{\delta^4\alpha^2}{\gamma_4^2\epec} \ergs
\label{sedec02max}
\eeq
is attained at
\beq
\nup{}_{ec,max} \approx \frac{1}{2}\nubr = 6.2\times 10^{19}\frac{\delta}{\epec} \us .
\label{sedec02fre}
\eeq
}

Now, we convert the results of equations (\ref{flunonlinG1}), (\ref{flunonlinG2}), and (\ref{flunonlinG3}) into SEDs, {which are the cases for $\alpha\gg 1$,} { yielding for the extreme Klein-Nishina-limit $\nubr<\nugbr$
\beq
\SED_{ec}(\nup{}) = 4.0\cdot 10^{34} R_{15} l_{ec}\frac{\delta^4\alpha^2}{\gamma_4^2\epec} \nonumber \\
\times \frac{\left( \frac{\nup{}}{\nubr} \right)^{1/2}}{\left( 1+\frac{\nup{}}{\nubr} \right)^{3/2} \left( 1+\frac{\nup{}}{\nugb} \right)^{2}} \left( 1-\frac{\nup{}}{\nucut} \right) \ergs ,
\label{sedec1}
\eeq
with  $\nugbr=1.02\cdot	10^{24}\delta\frac{\gamma_4}{\alpha} \us$, and $\nugb=1.23\cdot 10^{20}\delta\epgb \us$.} It peaks at
\be
\nup{}_{ec,max} \approx \frac{1}{2}\nubr = 6.2\cdot 10^{19} \delta \epec^{-1} \us
\label{sedec1fre}
\ee
{ with a maximum value of
\be
\SED_{ec,max} \approx 1.6\cdot 10^{34} R_{15} l_{ec} \frac{\delta^4\alpha^2}{\gamma_4^2\epec} \ergs .
\label{sedec1max}
\ee}
{ In the mild Klein-Nishina-limit $\nugbr<\nubr<\nugr$, with $\nugr=1.02\cdot 10^{24}\delta\gamma_4 \us$, we obtain
\beq
\SED_{ec}(\nup{}) = 4.0\cdot 10^{34} R_{15} l_{ec}\frac{\delta^4\alpha^2}{\gamma_4^2\epec} \nonumber \\
\times \frac{\left( \frac{\nup{}}{\nubr} \right)^{1/2}}{\left( 1+\frac{\nup{}}{\nugb} \right) \left( 1+\frac{\nup{}}{\nubr} \right)^{5/2}} \left( 1-\frac{\nup{}}{\nucut} \right) \ergs .
\label{sedec2}
\eeq
The maximum value
\be
\SED_{ec,max} \approx 2.0\cdot 10^{36} R_{15} l_{ec} \frac{\delta^4\alpha^{3/2}}{\gamma_4^{3/2}\epec^{1/2}} \ergs .
\label{sedec2max}
\ee}
is attained at
\be
\nup{}_{ec,max} \approx \nugb = 1.23\cdot 10^{24} \delta \frac{\gamma_4}{\alpha} \us
\label{sedec2fre}
\ee
{ Finally, we have the Thomson-limit $\nugr<\nubr$, where the SED becomes
\beq
\SED_{ec}(\nup{}) = 4.0\cdot 10^{34} R_{15} l_{ec} \frac{\delta^4\alpha^2}{\gamma_4^2\epec} \nonumber \\
\times \frac{\left( \frac{\nup{}}{\nubr} \right)^{1/2}}{\left( 1+\frac{\nup{}}{\nugb} \right)^{3/2} \left( 1+\frac{\nup{}}{\nugr} \right)^{2}} \left( 1-\frac{\nup{}}{\nucut} \right) \ergs .
\label{sedec3}
\eeq}
It peaks at
\be
\nup{}_{ec,max} \approx \nugb = 4.92\cdot 10^{28} \delta \frac{\gamma_4^2 \epec}{\alpha^2} \us
\label{sedec3fre}
\ee
(note that we use a different approximation for $\nugb$, that is for $\epgb$, compared to equation (\ref{sedec2fre})) { reaching a maximum value of
\be
\SED_{ec,max} \approx 1.9\cdot 10^{38} R_{15} l_{ec} \frac{\delta^4\alpha}{\gamma_4} \ergs .
\label{sedec3max}
\ee}
%
%
%
\subsection{Discussion}

\begin{figure*}
\begin{minipage}[t]{0.49\linewidth}
\centering \resizebox{\hsize}{!}
{\includegraphics{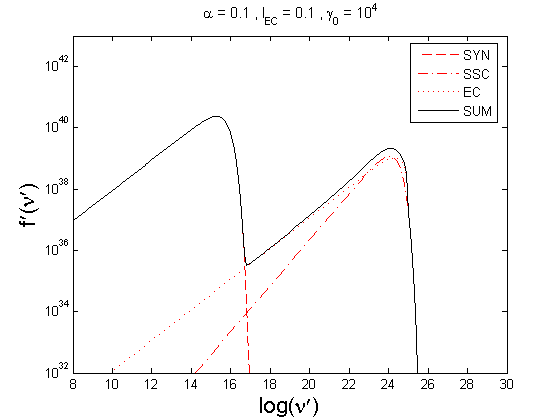}}
\end{minipage}
\hspace{\fill}
\begin{minipage}[t]{0.49\linewidth}
\centering \resizebox{\hsize}{!}
{\includegraphics{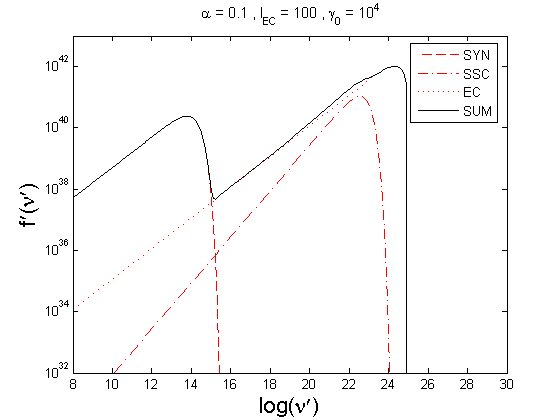}}
\end{minipage} \newline
\begin{minipage}[t]{0.49\linewidth}
\centering \resizebox{\hsize}{!}
{\includegraphics{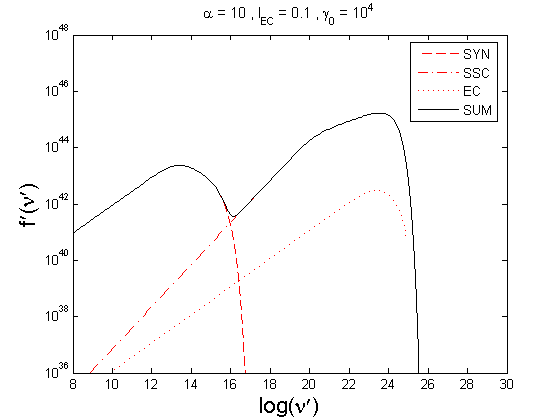}}
\end{minipage}
\hspace{\fill}
\begin{minipage}[t]{0.49\linewidth}
\centering \resizebox{\hsize}{!}
{\includegraphics{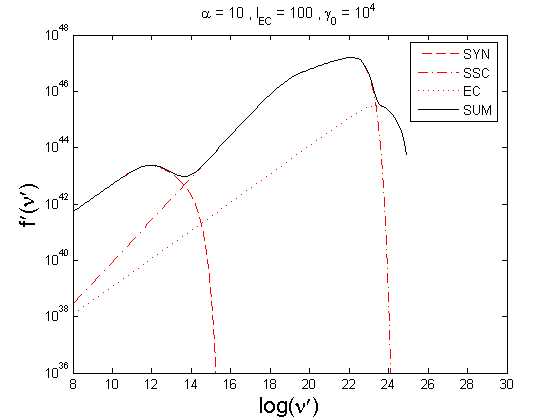}}
\end{minipage} \newline
\begin{minipage}[t]{0.49\linewidth}
\centering \resizebox{\hsize}{!}
{\includegraphics{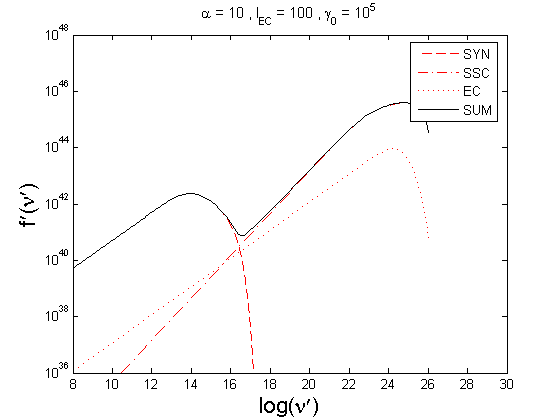}}
\end{minipage}
\hspace{\fill}
\begin{minipage}[t]{0.49\linewidth}
\centering \resizebox{\hsize}{!}
{\includegraphics{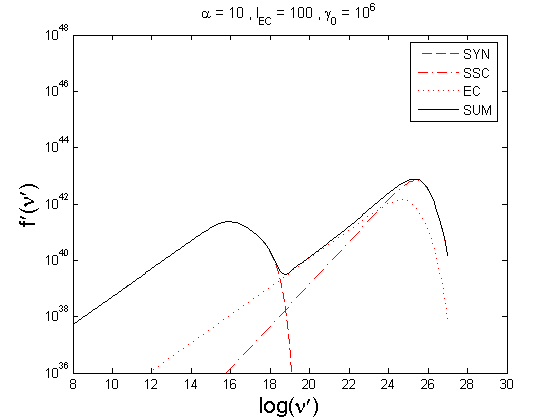}}
\end{minipage} 
\caption{Model SEDs for several cases of $\alpha$, $l_{ec}$, and $\gamma_0$. Specific parameters are given at the top of each plot, while the global parameters are given in the text. \textit{Top row:} $\alpha\ll 1$ in the Thomson limit for both cases of $l_{ec}$. \textit{Middle row:} $\alpha\gg 1$ in the Thomson limit for both cases of $l_{ec}$. \textit{Bottom row:} $\alpha\gg 1$ and $l_{ec}\gg 1$ in the mild and extreme Klein-Nishina regime, respectively.}
\label{fig1}
\end{figure*} 
In figure \ref{fig1} we present several model SED for different sets of parameters. Since the number of free parameters is fairly large a strict parameter study is beyond the scope of this paper, {although we will discuss some aspects below}. Therefore, we use some global parameters, which are unchanged through all figures, namely $\delta=10$, and $\Gamma_{b,1}=L^{\prime}_{46}=\tau_{-2}=R_{pc}^{\prime}=R_{15}=1$. The normalized external photon energy is also fixed at $\epec=10^{-4}$. This leaves us with three free parameters: the injection parameter $\alpha$, the relative strength of the external Compton cooling $l_{ec}$, and the initial electron Lorentz factor $\gamma_0$. These parameters are given at the top of the plots for each case. 

Since we use $l_{ec}$ as a free parameter and fixed almost all parameters it contains, we get the magnetic field strength $b$ as a dependent variable. Increasing $l_{ec}$ results in a decreasing magnetic field. This choice also demonstrates that the figures given as examples here only cover a very narrow range of possible realizations. 

{We note that for the top panels of figure \ref{fig1} we use equation (\ref{sedec0}) for the EC {component}, while we use equation (\ref{sedec3}) in the middle plots and equations (\ref{sedec2}), and (\ref{sedec1}) for the lower plots, respectively. This is due to our choice of parameters.}

{ The first obvious result is that for $\alpha\ll 1$ the inverse Compton component is significantly influenced, if not dominated by the EC component. For $\alpha\gg 1$ the main contribution comes from the SSC component, at least in the Thomson- and mild Klein-Nishina-limit. This is expected, since $\alpha$ controls the main cooling mechanism, which should also influence the SED.}

{ Similarly, this is also correct if one considers the linear coolings only, where the relative strength is controlled by $l_{ec}$. For example, in the top plots of figure \ref{fig1}, where we use equation (\ref{sedec0}), one can see that for $l_{ec}\ll 1$ the synchrotron component dominates the EC component, while for $l_{ec}\gg 1$ it is the other way around. Since $l_{ec}=|\dot{\gamma}_{ec}| / |\dot{\gamma}_{syn}|$, one could expect that the ratio of the peak values $R_{ec,s} = \SED_{ec,max} / \SED_{s,max} \propto l_{ec}$. This is certainly true for the Thomson limit of equation (\ref{sedec0}), as is shown in equation (\ref{Recs01}) of appendix \ref{sec:app4}. However, due to our choice of parameters the top panels of figure {\ref{fig1}} use the Klein-Nishina-limit of equation (\ref{sedec0}), where equation (\ref{Recs02}) demonstrates that the ratio is more involved and depends also on $\gamma_0$, and $\epec$. Using the parameters of the plots in the top row of figure \ref{fig1}, we find that the ratios become $R_{ec,s}\approx 0.07$ and $R_{ec,s}\approx 66$, respectively, consistent with the plots.}

In appendix \ref{sec:app4} we also show the ratios between the maximum values of the SSC and the EC components. As one can see, every $R_{ssc,ec} \propto \lec/l_{ec}$ with the consequence that the ratio is independent of $l_{ec}$ for $l_{ec}\gg 1$. Although the dependence on other parameters is rather involved and changes from case to case, we can say that a lower value of the external photon energy $\epec$ favors the EC component, while a higher value of the injection parameter $\alpha$ favors the SSC component, as one could expect.

On the other hand, if $l_{ec}\ll 1$, we find $R_{ssc,ec}\propto l_{ec}^{-1}$. This implies that the EC component is much reduced compared to the SSC component. This is also expected, since in this case the EC cooling is the weakest operating mechanism, if the other parameters are chosen accordingly. It is, of course, still possible that SSC could be lower than EC, since the ratios in appendix \ref{sec:app4} can always become smaller than unity.

{ As one can see in the lower panels of figure \ref{fig1}, we achieve similar results in the mild and extreme Klein-Nishina-regime, respectively. Especially in the latter case we have an example that the inclusion of the EC cooling has significant effects on the emerging spectrum, giving a different power-law and peak value compared to the case where either EC or SSC is neglected.}

As stated above, in figure \ref{fig1} we just show some model plots with only a very limited range of application, since we fixed most of the free parameters. Some of them, however, might have a significant impact on the resulting SED, like the radius $R$ of the plasma blob, which we set to $R_{15}=1$. This value corresponds to a light crossing time scale of $t_{lc}=R/(\delta c) \approx 0.9$h (with $\delta=10$), which is usually equaled with the variability time scale $t_{var}$ of the source. Naturally, a broad range of variability time scales is observed in blazars, ranging from minutes (as in the cases of PKS 2155-304, Aharonian et al. 2007, and PKS 1222+216, Tavecchio et al. 2011) to months (e.g. Abdo et al. 2010), indicating a large range of possible radii for the emission blobs. 

Increasing the radius of our plasma blob would result in a decreasing value of the injection parameter, since $\alpha\propto R^{-1}$ according to equation (\ref{alpha1}). This implies that the SSC component would become less important, leading to a dominating EC flux in the high energy regime.

On the other hand, long variability time scales can also be the result of inefficient cooling or other factors with long time scales, such as probably a rotating jet (e.g. Agudo et al. 2012). Our simple model assumes an instantaneous and effective cooling of the electrons far away from steady-state, favoring short variability time scales, at least if one is interested in the effects of the non-linear SSC cooling. Schlickeiser (2009), as well as Zacharias \& Schlickeiser (2010) showed that the non-linear process begins orders of magnitudes earlier to cool the electrons than the standard linear synchrotron cooling, implying shorter variability time scales. However, a true analyses of the variability should be performed with the help of light curves, which we aim to present in another paper.

We should also note that our discussion adopts a point-like emission zone, which means that we neglect additional effects from photon retardation variability processes from the finite size of the emission zone. For an extended analysis of these effects we refer to the discussion in Eichmann et al. (2010, 2012).

Another parameter, which might easily have different values, is the Doppler factor $\delta$. In the literature it usually covers a range $10-30$, but extreme values of $50$ and higher have been used to model blazar spectra (e.g. Begelman, Fabian \& Rees 2008), although the latter are mostly regarded as unrealistic. 

For our plots we used $\delta=10$. Increasing that value would not change the overall appearance of the SED, since in our simple model the dependencies on $\delta$ are the same in all cases, however the peak luminosities would increase significantly. In the plots in figure \ref{fig1} we achieve $\gamma$-ray luminosities roughly between $10^{39}\ergs$ and $10^{46}\ergs$. The former value is rather low for blazars. However, the values of the peak luminosity depend strongly on the parameters, which might differ from the ones used to create the plots.

As an example: With $\delta=50$, $\alpha=50$, and $l_{ec}=100$ one would arrive at a SSC luminosity of roughly $10^{50}\ergs$. This is close to the highest recorded luminosity of blazars, achieved by 3C 454.3 with a peak luminosity of $2.1\times 10^{50}\ergs$ (Abdo, et al. 2011b). In principle, the above stated values for $\alpha$ and $l_{ec}$ could be achieved by setting $N_{50}\approx 120$, and $b=0.14$. The high number of electrons in the source may not be that unrealistic, since a large outburst requires something extreme going on, such as a large mass load into the jet. According to equation (\ref{sedssc3fre}) the peak frequency of the SSC component would be at roughly $3\times 10^{23}\us$, which is reasonable (Abdo, et al. 2011b). Using equation (\ref{sedsyn2fre}) we obtain a synchrotron peak frequency roughly at $8\times 10^{11}\us$, which is in the far infrared regime. Still, a Doppler factor of $50$ is quite high, severely questioning the validity of the results of our simple model. Only a detailed spectral analysis, potentially also including a discussion of variability aspects (see above), can decide about the validity. However, such an analysis is beyond the scope of this paper.

\begin{table}[t]
	\caption{Energetic estimates of the plots in figure \ref{fig1}. \label{tab1}}
		\begin{tabular}{cccc}
			Plot & $u_e\; [\ergcm]$ & $P_e\; [\ergs]$ & $u_e/u_B$  \\
			\tableline
			TL & $9.1\times 10^{-4}$ & $9.2\times 10^{39}$ & $3.0\times 10^{-2}$ \\
			TR & $9.1\times 10^{-2}$ & $8.5\times 10^{41}$ & $2.4\times 10^{3}$ \\
			ML & $9.1$ & $8.9\times 10^{43}$ & $2.7\times 10^{2}$ \\
			MR & $9.1\times 10^{2}$ & $8.5\times 10^{45}$ & $2.5\times 10^{7}$ \\
			LL & $91$ & $8.5\times 10^{44}$ & $2.5\times 10^{6}$ \\
			LR & $9.1$ & $8.5\times 10^{43}$ & $2.5\times 10^{5}$ \\
		\end{tabular}
	\tablecomments{TL: top left, TR: top right, ML: middle left, \\ MR: middle right, LL: lower left, LR: lower right}
\end{table}

Another important aspect is the energy transported and dissipated by the content of the jet. The initial comoving energy density of the electrons is given by $u_e = q_0\gamma_0m_ec^2$. Using equation (\ref{q0eldens}), we obtain
\be
u_e = 0.09 \frac{\lec \alpha^2}{R_{15}\gamma_4} \ergcm .
\label{endensel1}
\ee
The power transported by these electrons is given by $P_e = \pi  R^2 c\beta_{\Gamma_b}\Gamma_b^2 u_e$, yielding with $\beta_{\Gamma_b}\approx 1$
\be
P_e = 8.4\times 10^{41} \Gamma_{b,1}^2 R_{15} \frac{\lec \alpha^2}{\gamma_4} \ergs .
\label{elpower1}
\ee
A widely used measure to quantify if the jet is dominated by particles or the magnetic field, is the equipartition parameter 
\be
\frac{u_e}{u_B} = 2.2 \frac{\lec \alpha^2}{b^2 R_{15}\gamma_4},
\label{equi1}
\ee
which can be rewritten with the help of equation (\ref{externalratio}), giving
\be
\frac{u_e}{u_B} = 24.4 \frac{R_{pc}^{\prime 2}}{L_{46}^{\prime}\tau_{-2}R_{15}} \frac{\lec l_{ec} \alpha^2}{\Gamma_{b,1}^2 \gamma_4} .
\ee

Table \ref{tab1} lists the values of these constraints for the plots given in figure \ref{fig1}. Depending on the used parameters we obtain a wide range of possible realizations.

A first result is that we are far away from equipartition with all our set-ups, and only the case with the lowest energies involved shows a dominating magnetic field. This is reasonable, since the number of electrons in this case is also at a minimum in our models. This is also the only case, where the synchrotron peak dominates the IC peak. However, we should note that this might be an artifact of our parameter settings, i.e. the inverse dependence of $b$ on $l_{ec}$, while we left all other parameters in $l_{ec}$ unchanged. Keeping $b$ at some value and changing some other parameters in $l_{ec}$ might give a different picture.

The energy density $u_e$ and the transferred power $P_e$ of the electron naturally increase with the number of particles in the blob. Comparing our values of $P_e$ with, e.g., Ghisellini et al. (2009), we find that our plots cover a range that is usually below obtained values for FSRQs, which is again due to our choice of parameters. The case with $\alpha=10$, $l_{ec}=100$, and $\gamma_4=1$ is, however, exactly in the range obtained by these authors. Higher values, as we used for the discussion about 3C 454.3 above, are also possible. So, from the energetics point of view our model is definitely supported.

%
\section{Summary and conclusions} \label{sec:con}
In this paper we aimed to extend the earlier work of Zacharias \& Schlickeiser (2012, ZS) regarding the importance of time-dependent synchrotron self-Compton cooling scenarios for modeling the spectral energy distributions of blazars. 

In ZS we were able to give an analytical SED for a pure synchrotron/SSC picture. However, especially in the context of flat spectrum radio quasars, a preference is given in the literature to external Compton scenarios, where the high energy component is modeled using external photon sources that are inverse Compton scattered by the jet electrons. We considered this effect here by introducing the external Compton cooling term into the differential equation describing the time- and energy-dependent behaviour of the electron distribution function. We used a rather simple model, where we assumed the external photons to be isotropically distributed in the frame of the blob. This is {justified by our intention to show in} an analytical demonstration the possibilities {and differences} of our approach {compared to the usual steady-state approach}. More {realistic} scenarios can mostly be treated only numerically, {since they contain more sophisticated assumptions and effects}. 

We then calculated the resulting external Compton intensity, the fluence and eventually the SED. Adapting the results of ZS regarding the synchrotron and the SSC SED, we have now obtained a complete analytical prediction of the appearance of blazar SEDs, if the electrons are cooled by the non-linear, time-dependent SSC channel in combination with the linear synchrotron and external Compton channels.

{The main addition to the conclusions of ZS is that it critically depends on the parameters of the source, which type of cooling dominates. For only weak EC cooling the results of ZS are recovered, while in the other case the shape of the high-energy component can change significantly. In appendix \ref{sec:app4} we calculated the ratios of the maximum values of each component with the other components, since they mark the dominance of one component over the others. These ratios especially show the importance of the choice of the specific values of the parameters, and demonstrate that a broad range of realizations is possible.

We also note that the nonlinear cooling has significant effects on the shape of the spectrum of all components leading to additional and also unique breaks, which strongly depend on $\alpha$. This highlights once more the severe effects of the time-dependent treatment.} 

To conclude, we can say that our investigation so far has clearly shown the importance of the inclusion of the time-dependent nature of the SSC cooling term, especially in blazar flaring states, which are far from steady-state. The resulting spectra differ significantly from the usual approach, where these effects are not taken into account. {Secondly, due to the broad range of free parameters in the EC model, the SSC component should not be dismissed beforehand while modeling blazars. As we were able to show here, the SSC component might be as important as the EC component, if the correct time-dependent treatment is applied, which is especially necessary in flaring states.}

In order to quantify {the differences between the time-dependent and steady-state approach} a bit further we intend to calculate the resulting light-curves and discuss the effects of optical thickness in a future work. The former might show in a clearer form that the SSC cooling actually acts a lot quicker than the usual cooling scenarios, while the latter might be important to constrain the free parameters, {such as the Doppler factor}.

%
%
\acknowledgements
We thank the anonymous referee for the constructive and detailed comments, which helped significantly improving the manuscript. \\
We acknowledge support from the German Ministry for Education and Research (BMBF) through Verbundforschung Astroteilchenphysik grant 05A11PC1 and the Deutsche Forschungsgemeinschaft through grant Schl 201/23-1. 

%
%
\appendix
\section{The external Compton emissivity} \label{sec:app1}

According to Dermer \& Schlickeiser (1993) the general emissivity due to external Compton scatterings is given by
\beq
j_{ec} = c\epss \intl_{0}^{\infty}\td{\epsilon}\oint\td{\Omega}\intl_{1}^{\infty}\td{\gamma}\oint\td{\Omega_e} (1-\beta_{\Gamma}\cos{\Psi}) \nonumber \\
\times \frac{u(\epsilon,\Omega)}{\epsilon}n(\gamma,\Omega_e)\sigma(\epss,\epsilon,\gamma,\Omega,\Omega_e) ,
\label{app:jec1}
\eeq
where $\Omega$, and $\Omega_e$ mark the direction of the incoming photon and the electron, respectively, and $\cos{\Psi}=\mu\mu_e+(1-\mu^2)^{1/2}(1-\mu_e^2)^{1/2}\cos{(\Phi-\Phi_e)}$ is the angle between the incident photon and the electron.

Assuming complete isotropy we can write $u(\epsilon,\Omega)=u(\epsilon)/4\pi$, $n(\gamma,\Omega_e)=n(\gamma)/4\pi$, and $\sigma(\epss,\epsilon,\gamma,\Omega,\Omega_e)=\sigma(\epss,\epsilon,\gamma)/4\pi$. Plugging this into equation (\ref{app:jec1}) the angle integrations can be easily performed, yielding
\beq
j_{ec} = \frac{c\epss}{4\pi}\intl_0^{\infty}\td{\epsilon}\frac{u(\epsilon)}{\epsilon}\intl_1^{\infty}\td{\gamma}n(\gamma)\sigma(\epss,\epsilon,\gamma),
\eeq
which agrees with equation (\ref{ecemi1}).

\section{Intensity and fluence for $\alpha\ll 1$} \label{sec:app2}

{
{ Using equation (\ref{nlinsol}) in equation (\ref{ecintens}) we find for the case of $\alpha\ll 1$:
\beq
I_{ec}(\epsilon_s,t) = \frac{3Rc\sigma_T}{16\pi}\epsilon_s\intl_0^{\infty}\td{\epsilon}\frac{\frac{4}{3}\Gamma_b^2 u^{\prime}_{ec}\DF{\epsilon-\epsilon_{ec}}}{\epsilon^2} \intl_{\gamma_{min}}^{\infty}\td{\gamma}\frac{q_0}{\gamma^2} G(q,\Gamma) \DF{\gamma-\frac{\gamma_0}{1+D_0\lec\gamma_0t}} ,
\label{linintens1}
\eeq
which leads directly to equation (\ref{linintens2}) with the substitution $\tau=D_0\lec\gamma_0t$ and the definition $I_0=Rc\sigma_Tq_0\Gamma_b^2u^{\prime}_{ec}/(4\pi\epsilon_{ec}^2\gamma_0^2)$.}

Integrating equation (\ref{linintens2}) with respect to time we yield
\beq
F(\epsilon_s) = F_0\epsilon_s \intl_0^{\frac{\gamma_0}{\gamma_{min}}-1}\td{\tau} (1+\tau)^2 G\left( q(\tau),\Gamma(\tau),\epsilon=\epsilon_{ec},\gamma=\frac{\gamma_0}{1+\tau} \right) ,
\label{flulin1}
\eeq
where we set $F_0 = I_0/(D_0\lec\gamma_0)$. \\
Inverting $q(\tau)$ gives
\be
\tau(q) = 2\epsilon_{ec}\gamma_0 q\left( \sqrt{1+\frac{1}{\epsilon_{s}\epsilon_{ec} q}}-1 \right)-1 .
\label{lintauq}
\ee
This can be substituted into equation (\ref{flulin1}) yielding
\beq
F(\epsilon_s) = 4F_0\epsilon_{ec}^3\gamma_0^3\epsilon_s \intl_{\frac{\epsilon_s}{4\epsilon_{ec}\gamma_0(\gamma_0-\epsilon_s)}}^{1}\td{q} \frac{q^2}{\sqrt{1+\frac{1}{\epsilon_s\epsilon_{ec}q}}} \left( \sqrteeq -1 \right)^4 G(q) .
\label{flulin2}
\eeq
With the above substitution the function $G$ simplifies to
\be
G(q) = G_0(q)+2\epss\epec q(1-q) .
\label{Gqsimp}
\ee
The final substitution $x:=(\epss\epec q)^{-1}$ gives
\beq
F(\epss) = 4F_0\gamma_0^3\epss^{-2} \intl_{(\epss\epec)^{-1}}^{\frac{4\gamma_0(\gamma_0-\epss)}{\epss^2}}\td{x} \frac{\left( \sqrt{1+x}-1 \right)^4}{x^4 \sqrt{1+x}} \left[ G_0\left( \frac{1}{\epss\epec x} \right) + \frac{2}{x}\left( 1-\frac{1}{\epss\epec x} \right) \right] .
\label{flulin3}
\eeq
This integral cannot be solved in a closed form. However, we can achieve approximative results giving reasonable solutions for most of the parameter space. 

For $\epss<\epec^{-1}$ one can see that both integration limits are much larger than unity, and so is $x$. Thus, we can approximate the fraction
\be
\frac{\left( \sqrt{1+x}-1 \right)^4}{x^4\sqrt{1+x}} \approx x^{-5/2} .
\ee
The bracket in equation (\ref{flulin3}) is to leading order in this approximation $G(x)\approx 1$. Now, this can be easily integrated, yielding
\beq
F(\epss<\epec^{-1}) \approx 4F_0\gamma_0^3\epss^{-2} \intl_{(\epss\epec)^{-1}}^{\frac{4\gamma_0(\gamma_0-\epss)}{\epss^2}}\td{x} x^{-5/2} \nonumber \\
= \frac{8}{3}F_0\gamma_0^3\epec^{3/2}\epss^{-1/2} \left[ 1-\left( \frac{\epss}{4\epec\gamma_0(\gamma_0-\epss)} \right)^{3/2} \right] .
\label{flulinsol1}
\eeq
For $\epss>\epec^{-1}$ and $\epss>2(\sqrt{2}-1)\gamma_0$ both integration limits are smaller than unity the fraction in the integral of equation (\ref{flulin3}) becomes
\be
\frac{\left( \sqrt{1+x}-1 \right)^4}{x^4\sqrt{1+x}} \approx 2^{-4} .
\ee
The leading order of $G(x)$ is here $G(x)\approx 2/(\epss\epec x^2)$. The integral, thus, results in
\beq
F(\epss>2(\sqrt{2}-1)\gamma_0) \approx 4F_0\gamma_0^3\epss^{-2} \intl_{(\epss\epec)^{-1}}^{\frac{4\gamma_0(\gamma_0-\epss)}{\epss^2}}\td{x} \frac{1}{2^3\epss\epec x^2} \nonumber \\
= \frac{1}{2}F_0\gamma_0^3\epss^{-2} \left[ 1- \frac{\epss}{4\epec\gamma_0(\gamma_0-\epss)} \right] .
\label{flulinsol2}
\eeq
For intermediate frequencies $\epec^{-1}<\epss<2(\sqrt{2}-1)\gamma_0$ we can split the integral into two regimes and use the above approximations:
\beq
F(\epec^{-1}<\epss<2(\sqrt{2}-1)\gamma_0) \approx 4F_0\gamma_0^3\epss^{-2} \left[ \intl_{(\epss\epec)^{-1}}^{1}\td{x} \frac{1}{2^3\epss\epec x^2} + \intl_{1}^{\frac{4\gamma_0(\gamma_0-\epss)}{\epss^2}}\td{x} x^{-5/2} \right] \nonumber \\
\approx \frac{19}{6}F_0\gamma_0^3\epss^{-2} .
\label{flulinsol3} 
\eeq	

Combining the above results we obtain the final expression for the external Compton fluence in the case of dominating linear cooling:
\be
F(\epss) = F_0\gamma_0^3\epec^{3/2} \frac{\epss^{-1/2}}{(1+\epec\epss)^{3/2}} \left( 1-\frac{\epss}{\ecut} \right) ,
\label{flulinsolgesapp}
\ee
where $\ecut = 4\epec\gamma_0^2/(1+4\epec\gamma_0)$ is the solution of the equation $\epss / (4\epec\gamma_0(\gamma_0-\epss)) = 1$, yielding equation (\ref{flulinsolges}).


\section{Intensity and fluence for $\alpha\gg 1$} \label{sec:app3}

\subsection{Early time limit, $t\leq t_c$}

{ For $t\leq t_c$ we use equation (\ref{nnonlinsol1}) in equation (\ref{ecintens}) and obtain
\beq
I_{ec}(\epsilon_s,t) = \frac{3Rc\sigma_T}{16\pi}\epsilon_s\intl_0^{\infty}\td{\epsilon}\frac{\frac{4}{3}\Gamma_b^2 u^{\prime}_{ec}\DF{\epsilon-\epsilon_{ec}}}{\epsilon^2} \intl_{\gamma_{min}}^{\infty}\td{\gamma}\frac{q_0}{\gamma^2}\HF{t_c-t} G(q,\Gamma) \DF{\gamma-\frac{\gamma_0}{(1+3D_0\lec\gamma_0\alpha^2t)^{1/3}}} .
\label{nonlinintens11}
\eeq }
Using the same definitions of $\tau$ and $I_0$ one can easily arrive at equation (\ref{nonlinintens12}). Then, the fluence integral can be written as
\beq
F(\epss) = F_0\epss\intl_0^{\infty}\td{\tau} (1+3\alpha^2\tau)^{2/3} G\left( q(\tau),\Gamma(\tau),\epsilon=\epsilon_{ec},\gamma=\frac{\gamma_0}{(1+3\alpha^2\tau)^{1/3}} \right) \HF{\tau_c-\tau} \HF{\frac{\gamma_0}{\gamma_{min}(\epec)}-(1+3\alpha^2\tau)^{1/3}} .
\label{flunnonlin1}
\eeq
Examining the Heaviside functions we find that $\tau_c$ is the upper limit as long as $\epss<\epgb$, while for $\epss>\epgb$ the upper limit becomes $\left[ \left( \frac{\gamma_0}{\gamma_{min}} \right)^3 - 1 \right]/3\alpha^2$. As long as they are not really needed we will refer to both of them as the respective variable with the subscript {\it up}. \\
The break energy is given by
\be
\epgb = \frac{\gamma_B}{1+\frac{1}{4\gamma_B\epec}} ,
\label{defepgb}
\ee
where $\gamma_B = \gamma_0/\alpha$. \\
Substituting $\mu=(1+3\alpha^2\tau)^{1/3}$ we obtain
\be
F(\epss) = \frac{F_0\epss}{\alpha^2} \intl_1^{\mu_{up}}\td{\mu} \mu^4 G(q,\Gamma,\mu) ,
\label{flunonlin2}
\ee
where $\mu_{up}(\epss<\epgb) = \alpha$, and $\mu_{up}(\epss>\epgb) = \gamma_0/\gamma_{min}$. \\
Inverting $q(\mu)$ yields
\be
\mu = 2\epec\gamma_0 q \left( \sqrteeq - 1 \right) .
\label{nonlinmuq}
\ee
Inserting this into equation (\ref{flunonlin2}) with $q_{up}(\epss<\epgb) = \epss\alpha^2/(4\epec\gamma_0(\gamma_0-\epss\alpha))$, and $q_{up}(\epss>\epgb) = 1$, we find
\beq
F(\epss) = \frac{16F_0\epec^5\gamma_0^5}{\alpha^2}\epss \intl_{\frac{\epss}{4\epec\gamma_0(\gamma_0-\epss)}}^{q_{up}}\td{q} \frac{q^4}{\sqrteeq} \left( \sqrteeq - 1 \right)^6 \left[ G_0(q)+2\epss\epec q(1-q) \right] .
\label{flunonlin3} 
\eeq
Using again the substitution $x=(\epss\epec q)^{-1}$ we eventually obtain
\beq
F(\epss) = \frac{16F_0\gamma_0^5}{\alpha^2}\epss^{-4} \intl_{x_{down}}^{\frac{4\gamma_0(\gamma_0-\epss)}{\epss^2}}\td{x} \frac{\left( \sqrt{1+x} - 1 \right)^6}{x^6 \sqrt{1+x}} \left[ G_0\left( \frac{1}{\epss\epec x} \right) + \frac{2}{x}\left( 1-\frac{1}{\epss\epec x} \right) \right] ,
\label{flunonlin4}
\eeq
with the lower limits $x_{down}(\epss<\epgb) = 4\gamma_0(\gamma_0-\alpha\epss)/\epss^2\alpha^2$, and $x_{down}(\epss>\epgb) = (\epss\epec)^{-1}$. \\
Since this integral cannot be solved in a closed form, as well, we will now consider similar approximations as in the linear case. Additionally, we will now consider each case of the lower limit in turn. \\

Beginning with $\epgb<\epss<\ecut$ we use the lower limit $x_{down}(\epss>\epgb) = (\epss\epec)^{-1}$. In case of $\epec>(2(\sqrt{2}-1)\gamma_B)^{-1}$ we find for $2(\sqrt{2}-1)\gamma_0<\epss<\ecut$ that both limits are lower than unity and we approximate the fraction in equation (\ref{flunonlin4}) with
\be
\frac{\left( \sqrt{1+x} - 1 \right)^6}{x^6 \sqrt{1+x}} \approx 2^{-6} ,
\ee
and $G(x)$ becomes to leading order $G(x)\approx 2/(\epss\epec x^2)$. Hence,
\beq
F(2(\sqrt{2}-1)\gamma_0<\epss<\ecut) \approx \frac{16F_0\gamma_0^5}{\alpha^2}\epss^{-4} \intl_{(\epss\epec)^{-1}}^{\frac{4\gamma_0(\gamma_0-\epss)}{\epss^2}}\td{x} \frac{1}{2^5\epss\epec x^2} \nonumber \\
= \frac{F_0\gamma_0^5}{2\alpha^2}\epss^{-4} \left[ 1-\frac{\epss}{\ecut} \right] .
\label{flunonlin5a}
\eeq
For $\epgb<\epss<\sqrttwo\gamma_0$ we see that the upper limit is larger than unity, while the lower limit is still smaller than unity. We will, therefore, split the integral. For the part below unity we use the above mentioned approximation, while for the part larger than unity we approximate the integrand with $x^{-7/2}$. Thus,
\beq
F(\epgb<\epss<\sqrttwo\gamma_0) \approx \frac{16F_0\gamma_0^5}{\alpha^2}\epss^{-4} \left[ \intl_{(\epss\epec)^{-1}}^{1} \frac{\td{x}}{2^5\epec\epss x^2} + \intl_{1}^{\frac{4\gamma_0(\gamma_0-\epss)}{\epss^2}} \td{x} x^{-7/2} \right] \nonumber \\
\approx \frac{F_0\gamma_0^5}{2\alpha^2}\epss^{-4} .
\label{flunonlin5b}
\eeq
Combining the last two equation we obtain
\be
F(\epgb<\epss<\ecut) = \frac{F_0\gamma_0^5}{2\alpha^2} \epss^{-4} \left( 1-\frac{\epss}{\ecut} \right) .
\label{flunonlin5}
\ee

For $\epec<(\sqrttwo\gamma_B)^{-1}$ we can make use of equations (\ref{flunonlin5a}) and (\ref{flunonlin5b}), while $x\ll 1$ and $x\approx 1$, respectively, yielding
\beq
F(\sqrttwo\gamma_0<\epss<\ecut) \approx  \frac{F_0\gamma_0^5}{2\alpha^2}\epss^{-4} \left[ 1-\frac{\epss}{\ecut} \right] ,
\label{flunonlin6a} 
\eeq
and
\beq
F(\epec^{-1}<\epss<\sqrttwo\gamma_0) \approx \frac{F_0\gamma_0^5}{2\alpha^2}\epss^{-4} .
\label{flunonlin6b}
\eeq
If both limits are larger than unity we approximate the integrand with $x^{-7/2}$ and obtain
\beq
F(\epgb<\epss<\epec^{-1}) \approx \frac{16F_0\gamma_0^5}{\alpha^2}\epss^{-4} \intl_{(\epss\epec)^{-1}}^{\frac{4\gamma_0(\gamma_0-\epss)}{\epss^2}}\td{x} x^{-7/2} \nonumber \\
= \frac{32F_0\gamma_0^5\epec^{5/2}}{5\alpha^2}\epss^{-3/2}\left[ 1-\left( \frac{\epss}{\ecut} \right)^{5/2} \right] .
\label{flunonlin6c}
\eeq
Combining the equations (\ref{flunonlin6a}) - (\ref{flunonlin6c}) we can write the final result as
\beq
F(\epgb<\epss<\ecut) =  \frac{F_0\gamma_0^5\epec^{5/2}}{2\alpha^2} \frac{\epss^{-3/2}}{(1+\epec\epss)^{5/2}} \left( 1-\frac{\epss}{\ecut} \right) .
\label{flunonlin6}
\eeq

Now, we turn our attention to the case where $\epss<\epgb$, which means that the lower limit in the integral of equation (\ref{flunonlin4}) becomes $x_{down} = 4\gamma_0(\gamma_0-\alpha\epss) / \epss^2\alpha^2$. \\
The first thing, we notice, is that the upper limit is always larger than unity. This is also true for the lower limit, except for $\epec^{-1}<\sqrttwo\gamma_B$. In the latter case, we obtain with the same approximations as above:
\beq
F(\sqrttwo\gamma_B<\epss<\epgb) \approx \frac{16F_0\gamma_0^5}{\alpha^2}\epss^{-4} \left[ \intl_{\frac{4\gamma_0(\gamma_0-\alpha\epss)}{\epss^2\alpha^2}}^{1} \frac{\td{x}}{2^5\epec\epss x^2} + \intl_{1}^{\frac{4\gamma_0(\gamma_0-\epss)}{\epss^2}} \td{x} x^{-7/2} \right] \nonumber \\
\approx \frac{F_0\gamma_0^3}{8\epec}\epss^{-3} .
\label{flunonlin7a}
\eeq
In the case, where both limits are larger than unity, we find
\beq
F(\epss<\epgb) \approx \frac{16F_0\gamma_0^5}{\alpha^2}\epss^{-4} \intl_{\frac{4\gamma_0(\gamma_0-\alpha\epss)}{\epss^2\alpha^2}}^{\frac{4\gamma_0(\gamma_0-\epss)}{\epss^2}}\td{x} x^{-7/2} \nonumber \\
\approx \frac{F_0\alpha^3}{5}\epss .
\label{flunonlin7b}
\eeq

Finally, we can give the fluence in the early time limit covering all energies. However, we have to distinguish three cases for the external photon energy. \\
For $\epec^{-1}<\sqrttwo\gamma_B$ we have
\beq
F(\epss) = \frac{F_0\alpha^3}{5} \frac{\epss}{\left( 1+\frac{\epss}{\epgb} \right)\left( 1+\frac{\epss}{\sqrttwo\gamma_B} \right)^{4}} \left( 1-\frac{\epss}{\ecut} \right) ,
\label{flunonling1app}
\eeq
while for $\sqrttwo\gamma_B<\epec^{-1}<\sqrttwo\gamma_0$ we get
\beq
F(\epss) = \frac{F_0\alpha^3}{5} \frac{\epss}{\left( 1+\frac{\epss}{\epgb} \right)^{5/2} \left( 1+\epec\epss \right)^{5/2}} \left( 1-\frac{\epss}{\ecut} \right) . 
\label{flunonling2app}
\eeq
The last part is for $\sqrttwo\gamma_0<\epec^{-1}$ and becomes
\beq
F(\epss) = \frac{F_0\alpha^3}{5} \frac{\epss}{\left( 1+\frac{\epss}{\epgb} \right)^{5/2} \left( 1+\frac{\epss}{\sqrttwo\gamma_B} \right)^{5/2}} \left( 1-\frac{\epss}{\ecut} \right) . 
\label{flunonling3app}
\eeq
Hence, we obtained equations (\ref{flunonling1}) - (\ref{flunonling3}).


\subsection{Late time limit, $t> t_c$}

{ As before, we can find the intensity for $t\geq t_c$, if we insert equation (\ref{nnonlinsol2}) into equation (\ref{ecintens}):
\beq
I_{ec}(\epsilon_s,t) = \frac{3Rc\sigma_T}{16\pi}\epsilon_s\intl_0^{\infty}\td{\epsilon}\frac{\frac{4}{3}\Gamma_b^2 u^{\prime}_{ec}\DF{\epsilon-\epsilon_{ec}}}{\epsilon^2} \intl_{\gamma_{min}}^{\infty}\td{\gamma}\frac{q_0}{\gamma^2}\HF{t-t_c} G(q,\Gamma) \DF{\gamma-\frac{\gamma_0}{\frac{1+2\alpha^3}{3\alpha^2}+D_0\lec\gamma_0t}} .
\label{nonlinintens21}
\eeq}
Equation (\ref{nonlinintens22}) is recovered if one uses the definitions of $\tau$ and $I_0$, again. The fluence then follows to be
\beq
F(\epss) = F_0\epss \intl_{\tau_c}^{\frac{\gamma_0}{\gamma_{min}}-\frac{1+2\alpha^3}{3\alpha^2}} \td{\tau} \left( \frac{1+2\alpha^3}{3\alpha^2}+\tau \right)^2 G\left( q,\Gamma,\epsilon=\epec,\gamma=\frac{\gamma_0}{\frac{1+2\alpha^3}{3\alpha^2}+\tau} \right) .
\label{flunonlin8}
\eeq
Substituting $y=\frac{1+2\alpha^3}{3\alpha^2}+\tau$ we obtain
\beq
F(\epss) = F_0\epss \intl_{\alpha}^{\frac{\gamma_0}{\gamma_{min}}}\td{y} y^2 G\left( q,\Gamma,\epsilon=\epec,\gamma=\frac{\gamma_0}{y} \right)
\label{flunonlin9} .
\eeq
Inverting $q(y)$ gives $y=2\epec\gamma_0 q \left( \sqrteeq - 1 \right)$, which leads to
\beq
F(\epss) = 8F_0\epec^3\gamma_0^3\epss \intl_{\frac{\epss\alpha^2}{4\epec\gamma_0(\gamma_0-\alpha\epss)}}^{1} \td{q} \frac{q^2\left( \sqrteeq -1 \right)^4}{\sqrteeq} \left[ G_0(q) + 2\epss\epec q(1-q) \right] .
\label{flunonlin10}
\eeq
Finally, we use again $x=(\epec\epss q)^{-1}$ and, thus, the fluence becomes
\beq
F(\epss) = 8F_0\gamma_0^3\epss^{-2} \intl_{(\epss\epec)^{-1}}^{\frac{4\gamma_0(\gamma_0-\alpha\epss)}{\epss^2\alpha^2}} \td{x} \frac{\left( \sqrt{1+x}-1 \right)^4}{x^4\sqrt{1+x}} \left[ G_0\left( \frac{1}{\epss\epec x} \right) + \frac{2}{x}\left( 1-\frac{1}{\epss\epec x} \right) \right] .
\label{flunonlin11}
\eeq

As before, we will consider approximative results. For $\epec^{-1}<\sqrttwo\gamma_B<\epss<\epgb$ both limits are smaller than unity, and we approximate the integrand with $(2^3\epss\epec x^2)^{-1}$ giving
\beq
F(\sqrttwo\gamma_B<\epss<\epgb) \approx 8F_0\gamma_0^3\epss^{-2} \intl_{(\epss\epec)^{-1}}^{\frac{4\gamma_0(\gamma_0-\alpha\epss)}{\epss^2\alpha^2}} \td{x} \frac{1}{2^3\epss\epec x^2} \nonumber \\
= F_0\gamma_0^3\epss^{-2} \left( 1-\frac{\epss}{\epgb} \right) .
\label{flunonlin12a}
\eeq
For $\epss<\epec^{-1}<\sqrttwo\gamma_B<\epgb$ both limits are larger than unity and the integrand can be written as $x^{-5/2}$, yielding
\beq
F(\epss<\epec^{-1}<\sqrttwo\gamma_B) \approx 8F_0\gamma_0^3\epss^{-2} \intl_{(\epss\epec)^{-1}}^{\frac{4\gamma_0(\gamma_0-\alpha\epss)}{\epss^2\alpha^2}} \td{x} x^{-5/2} \nonumber \\
= \frac{16}{3}F_0\epec^{3/2}\gamma_0^3\epss^{-1/2} \left[ 1-\left( \frac{\epss}{\epgb} \right)^{3/2} \right] .
\label{flunonlin12b}
\eeq
In the intermediate part $\epec^{-1}<\epss<\sqrttwo\gamma_B<\epgb$ we can use both limits to obtain
\beq
F(\epec^{-1}<\epss<\sqrttwo\gamma_B) \approx 8F_0\gamma_0^3\epss^{-2} \left[ \intl_{(\epss\epec)^{-1}}^{1} \td{x} \frac{1}{2^3\epss\epec x^2} + \intl_{1}^{\frac{4\gamma_0(\gamma_0-\alpha\epss)}{\epss^2\alpha^2}} \td{x} x^{-5/2} \right] \nonumber \\
\approx F_0\gamma_0^3 \epss^{-2} .
\label{flunonlin12c}
\eeq

Collecting terms we find the fluence in the late time limit for $\epec^{-1}<\sqrttwo\gamma_B$
\beq
F(\epss<\epgb) = F_0\gamma_0^3\epec^{3/2} \frac{\epss^{-1/2}}{(1+\epec\epss)^{3/2}} \left( 1-\frac{\epss}{\epgb} \right) .
\label{flunonling4app}
\eeq
For $\epec^{-1}>\sqrttwo\gamma_B$ we obtain a single power-law in the form
\beq
F(\epss<\epgb) = \frac{16}{3}F_0\epec^{3/2}\gamma_0^3\epss^{-1/2} \left( 1-\frac{\epss}{\epgb} \right) .
\label{flunonling5app}
\eeq
Both equations equal equations (\ref{flunonling4}) and (\ref{flunonling5}), respectively.
}


\section{The ratios of the maximum values} \label{sec:app4}

{
The ratios of the maximum or peak values are defined as
\beq
R_{i,j} = \frac{\SED_{i,max}}{\SED_{j,max}} .
\label{Rall}
\eeq


\subsection{External to synchrotron}

{ In the case $\alpha\ll 1$ we use equations (\ref{sedsyn1max}) and (\ref{sedec01max}) for the Thomson-limit of the EC component, and obtain
\beq
R_{ec,s} = 1.8\; l_{ec} .
\label{Recs01}
\eeq
In the Klein-Nishina-limit of the EC component we use equation (\ref{sedec02max}), and get
\beq
R_{ec,s} = 6.6\times 10^{-5} \frac{l_{ec}}{\gamma_4\epec} .
\label{Recs02}
\eeq

For $\alpha\gg 1$ we have to distinguish between the three cases  for $\nubr$. For the extreme Klein-Nishina-limit we use equations (\ref{sedsyn2max}) and (\ref{sedec1max}), giving
\beq
R_{ec,s} = 6.6\times 10^{-5} l_{ec} \frac{\alpha}{\gamma_4\epec} .
\label{Recs1}
\eeq
For the mild Klein-Nishina-limit we use equation (\ref{sedec2max}), yielding
\beq
R_{ec,s} = 8.3\times 10^{-3} l_{ec} \left( \frac{\alpha}{\gamma_4\epec} \right)^{1/2} ,
\label{Recs2}
\eeq
while we use equation (\ref{sedec3max}) for the Thomson-limit, resulting in
\beq
R_{ec,s} = 0.8\; l_{ec} .
\label{Recs3}
\eeq}


\subsection{SSC to synchrotron}

In this case we have to distinguish not only between the possible values of $\alpha$, but also between the different cases of the Klein-Nishina-parameter $K$.

For $\alpha\ll 1$ we use for the synchrotron peak value equation (\ref{sedsyn1max}), again, while for SSC peak value for $K\ll 1$ we use equation (\ref{sedssc1max}). The ratio then becomes
\beq
R_{ssc,s} = 4.4 \lec \alpha^2 .
\label{Rsscs1}
\eeq
For $K\gg 1$ we use equation (\ref{sedssc2max}), giving
\beq
R_{ssc,s} = 26.8 \lec \frac{\alpha^2}{b\gamma_4^3} .
\label{Rsscs2}
\eeq

Using equation (\ref{sedsyn2max}) for the synchrotron peak value in the case $\alpha\gg 1$, we find with equation (\ref{sedssc3max}) the ratio in the case\footnote{Note the printing error in ZS. The correct value (apart from $\lec$) is given here.} $K\ll 1$:
\beq
R_{ssc,s} = 6.8 \lec \alpha .
\label{Rsscs3}
\eeq
The ratio for the case $1\ll K\ll \alpha^3$ can be found by using equation (\ref{sedssc4max}), yielding
\beq
R_{ssc,s} = 5.2 \lec \frac{\alpha}{b^{1/3} \gamma_4} .
\label{Rsscs4}
\eeq
Lastly, we obtain the ratio for $1\ll \alpha^3\ll K$ with the help of equation (\ref{sedssc5max}):
\beq
R_{ssc,s} = 14.4 \lec \frac{\alpha^3}{b\gamma_4^3} .
\label{Rsscs5}
\eeq

Compared to ZS all ratios gain a factor $\lec$, raising the ratio, i.e. the Compton dominance, potentially by a lot.


\subsection{SSC to external}

Although this ratio is not between the two different components of the blazar SED, it still might be useful to give the ratios between the possible realizations of the high-energy component, since, depending on the parameters, it might be possible to discriminate between either the SSC or the external Compton scenario.  

\subsubsection{Small injection parameter, $\alpha\ll 1$}

{ Firstly, we use the Thomson-limit for the external Compton SED, where the peak value is given by equation (\ref{sedec01max}). Equation (\ref{sedssc1max}) gives the case $K\ll 1$ for the SSC component, yielding the ratio 
\beq
R_{ssc,ec} = 2.3 \frac{\lec}{l_{ec}} \alpha^2 .
\label{Rsscec1}
\eeq
For $K\gg 1$ we obtain with equation (\ref{sedssc2max})
\beq
R_{ssc,ec} = 14.3 \frac{\lec}{l_{ec}} \frac{\alpha^2}{b\gamma_4^3} .
\label{Rsscec2}
\eeq
Secondly, we use equation (\ref{sedec02max}) for the peak value of the external Compton SED in the Klein-Nishina-limit. For $K\ll 1$ we obtain with equation (\ref{sedssc1max}):
\beq
R_{ssc,ec} = 6.3\times 10^4 \frac{\lec}{l_{ec}} \alpha^2 \gamma_4 \epec .
\label{Rsscec3}
\eeq
For $K\gg 1$ we use equation (\ref{sedssc2max}), yielding
\beq
R_{ssc,ec} = 4.0\times 10^5 \frac{\lec}{l_{ec}} \frac{\alpha^2\epec}{b\gamma_4^2} .
\label{Rsscec4}
\eeq}

\subsubsection{Large injection parameter, $\alpha\gg 1$}

In this case we obtain nine different ratios, which will be given below.

{ Beginning with the case $K\ll 1$ we will use equation (\ref{sedssc3max}) for the SSC peak values. For $\nubr<\nugbr$ we use equation (\ref{sedec1max}), which becomes
\beq
R_{ssc,ec} = 1.0\times 10^5 \frac{\lec}{l_{ec}} \gamma_4\epec .
\label{Rsscec1a}
\eeq
In the case $\nugbr<\nubr<\nugr$ we take equation (\ref{sedec2max}), yielding
\beq
R_{ssc,ec} = 8.0\times 10^2 \frac{\lec}{l_{ec}} \left( \alpha\gamma_4\epec \right)^{1/2} .
\label{Rsscec1b}
\eeq
Thirdly, we have the case $\nugr<\nubr$, where we use equation (\ref{sedec3max}) to obtain
\beq
R_{ssc,ec} = 8.4 \frac{\lec}{l_{ec}} \alpha .
\label{Rsscec1c} 
\eeq

Continuing with the case $1\ll K\ll \alpha^3$ we take equation (\ref{sedssc4max}) for the SSC peak values. With equation (\ref{sedec1max}) we get for $\nubr<\nugbr$
\beq
R_{ssc,ec} = 7.5\times 10^4 \frac{\lec}{l_{ec}} \frac{\epec}{b^{1/3}} .
\label{Rsscec2a}
\eeq
For $\nugbr<\nubr<\nugr$ we use equation (\ref{sedec2max}), again, giving
\beq
R_{ssc,ec} = 6.0\times 10^2 \frac{\lec}{l_{ec}} \frac{\alpha^{1/2}\epec^{1/2}}{b^{1/3}\gamma_4^{1/2}} .
\label{Rsscec2b}
\eeq
In the case $\nugr<\nubr$ equation (\ref{sedec3max}) gives the correct ratio:
\beq
R_{ssc,ec} = 6.3 \frac{\lec}{l_{ec}} \frac{\alpha}{b^{1/3}\gamma_4} .
\label{Rsscec2c} 
\eeq

Finally, we use equation (\ref{sedssc5max}) for the case $1\ll \alpha^3\ll K$. With equation (\ref{sedec1max}) we yield for $\nubr<\nugbr$
\beq
R_{ssc,ec} = 2.1\times 10^5 \frac{\lec}{l_{ec}} \frac{\alpha^2\epec}{b\gamma_4^2} .
\label{Rsscec3a}
\eeq
In the case $\nugbr<\nubr<\nugr$ we take equation (\ref{sedec2max}), obtaining
\beq
R_{ssc,ec} = 1.7\times 10^3 \frac{\lec}{l_{ec}} \frac{\alpha^{5/2}\epec^{1/2}}{b\gamma_4^{5/2}} .
\label{Rsscec3b}
\eeq
Lastly, we have the case $\nugr<\nubr$, where we find with equation (\ref{sedec3max}) 
\beq
R_{ssc,ec} = 17.9 \frac{\lec}{l_{ec}} \frac{\alpha^3}{b\gamma_4^3} .
\label{Rsscec3c} 
\eeq}

We see in all cases that for $l_{ec}\gg 1$ the ratio is independent of this value, while for $l_{ec}\ll 1$ the ratio becomes $R_{ssc,ec}\propto l_{ec}^{-1}$. The latter implies a growing dominance of the SSC component over the EC component, which is expected. The former shows that it critically depends on the other parameters, which component might dominate the inverse Compton hump.

}

%
%

%
\end{document}